\documentstyle[epsfig,amsmath,amssymb,graphicx,soul,color,deluxetable,subfigure,
lscape]{jaa}

\def\gsim{\;\rlap{\lower 2.5pt\hbox{$\sim$}}\raise 1.5pt\hbox{$>$}\;}
\def\lsim{\;\rlap{\lower 2.5pt\hbo time lag between the formation x{$\sim$}}\rai
se 1.5pt\hbox{$<$}\;}
\def\la{\mathrel{\hbox{\rlap{\hbox{\lower4pt\hbox{$\sim$}}}\hbox{$<$}}}}
\def\ga{\mathrel{\hbox{\rlap{\hbox{\lower4pt\hbox{$\sim$}}}\hbox{$>$}}}}

\begin{document}
\title[Asteroseismology from ARIES]{Asteroseismology of Pulsating Stars}
\author[S. Joshi \& Y. C. Joshi]{Santosh Joshi \& Yogesh C. Joshi\\
\\
Aryabhatta Research Institute of Observational Sciences(ARIES), Manora Peak, Nainital 263 002, India\\
}

\pubyear{xxxx}
\volume{xx}
\date{Received xxx; accepted xxx}
\maketitle
\label{firstpage}
\begin{abstract}

\end{abstract}
The success of helioseismology is due to its capability of measuring $p$-mode 
oscillations in the Sun.  This allows us to extract informations on the internal 
structure and rotation of the Sun from the surface to the core. Similarly, asteroseismology 
is the study of the internal structure of the stars as derived from stellar
oscillations.  In this review we highlight the progress in the  observational
asteroseismology, including some basic theoretical aspects.  In particular,
we discuss our contributions to asteroseismology through the study of chemically peculiar stars 
under the ``Nainital-Cape Survey'' project being conducted at ARIES, Nainital since 1999.  This survey aims to detect new rapidly-pulsating Ap (roAp) stars in the northern 
hemisphere. We also discuss the contribution of ARIES towards the asteroseismic 
study of the compact pulsating variables.  We comment on the future prospects of 
our project in the light of the new  optical 3.6-m telescope to be install at Devasthal (ARIES). Finally, we present a preliminary optical design of the high-speed imaging photometers for this telescope.  


\section{Introduction} \label{sec:st_int}

The conference entitled ``Plasma Processes in Solar and Space Plasma at Diverse 
Spatio-Temporal Scales : Upcoming Challenges in the Science and Instrumentation'' 
took place during 26-28 March 2014 when the preparation for the installation of the 3.6-m telescope at Devasthal (ARIES) is going on and the state-of-art instruments would soon be available for the study of faint astronomical objects with high-temporal resolution. Although this conference was devoted to the study of the solar interior (helioseismology) but asteroseismology was one of the chosen topic for this conference. In the following sub-sections we give a brief introduction to the subject followed by the techniques.

\subsection{Variable Stars} 

Many stars show light variations on a time-scale shorter than their evolutionary 
changes and such objects are known as variable stars. Two major groups of variable stars are known:  extrinsic and intrinsic variables. The light variations in extrinsic variables are due to external factors. For example, the light variations in eclipsing binary stars is caused by two stars passing in front of each other, so that light coming from one of them is periodically blocked by another one. By analysing the light curves of eclipsing binaries, one can estimate the fundamental parameters such as their masses, radius etc. On the other hand, the light variations in intrinsic variables arise in the stars themselves.  For example, as the size and shape of a star changes due to pulsations. The Sun is a variable star whose magnetic activity varies on a time-scale of approximately
11 years. However, significant variation of a period around five minutes is also observed in Sun. These short period oscillations have very low-amplitude and are only detectable due to it's proximity.  

The variable stars have been used to estimate various basic physical 
stellar parameters ever since the discovery of Mira by David Fabricius (Olbers 1850). 
The 11-months periodic brightness of Mira shows a 10-mag difference between minimum and maximum brightness. Shapley (1914) suggested that variability among Cepheids and cluster 
variables are  due to internal or surface pulsations
in the star itself. Pigott (1785) discovered the first pulsating $\delta$-Cephei 
(Cepheid) star and the  period-luminosity (P-L) relation  was discovered  by 
Henrietta Swan Leavitt (Leavitt \& Pickering 1912) which is the foundation of the cosmic distance scale.  

The photometric and spectroscopic observational techniques with modest-sized telescopes 
of diameters 1 to 4-m  are well suited for the detection and study of stellar 
variability. The photometric precision attained with such
telescopes is normally of the order of few milli-magnitudes (mmag).  Here, the limiting factor is the instability in the atmosphere caused by scintillation and extinction.
The radial velocity precision attained by spectroscopy could be of the
order of a cm/sec. However, the atmospheric effects does not play major role in spectroscopic observations, hence pulsations are more easily detected in comparison to the photometric observations.
The prospects for high-precision photometry and spectroscopy in the field of asteroseismology 
are thus very promising.  For this reason we are developing a high-resolution spectrograph and time-series CCD photometers for the up-coming 3.6-m telescope at Devasthal (ARIES). 

\subsection{Helioseismology}

The Sun is an outstanding example of the stellar seismology. The five-minutes oscillatory motion in the atmosphere of the Sun was detected in the early 1960's by Leighton et al. (1962). The oscillations observed at the surface of the Sun manifest themselves in small motions 
similar to the seismic waves generated at the surface of Earth during
earthquakes. This discovery gave birth to the helio-seismology which has proved to be an extremely successful technique in probing the physics and dynamics of the solar interior (Christensen-Dalsgaard 2002; Basu \& Antia 2008; Chaplin \& Basu 2008). A large number of oscillation modes ($\sim10^7$) are thought to be simultaneously excited in Sun and each mode carries information on each part of the solar interior. When averaged over the solar disk, the perturbations due to these modes average to zero, rendering them undetectable.
These studies have contributed greatly to a clearer understanding of the 
Sun by measuring the characteristics of the $p$-mode oscillation spectrum. 

Solar oscillations are the result of the excitation of global modes by
convective motion (forced oscillations) where pressure is the restoring
force ($p$-modes). On the surface of Sun, these oscillations are observed as Doppler shifts of spectrum lines with amplitudes of individual modes not exceeding 0.1 m/sec or as whole-disk luminosity variations with amplitudes not exceeding a few part per million (ppm). The observational confirmation of the five-minutes oscillations was done by Deubner (1975) and Rhodes et al. (1977). Ulrich (1970) and Leibacher \& Stein (1971) proposed  that the waves observed on the surface of Sun are standing-wave in nature. The periods of solar-oscillations are in the range of 3--15 min with peak amplitudes around 5-min (Fletcher et al. 2010). A number of studies have also been claimed  the  detection of $g$-modes in the Sun (Garc{\'i}a et al. 2008; Appourchaux  et al. 2010)) but more studies are required for their confirmation. This proceedings contain many articles on helioseismology where further details can be found.

\subsection{From Helioseismology to Asteroseismology}

 During the last decades, many observational and theoretical efforts in the study of the 
acoustic modes of solar oscillations have brought to a detailed knowledge of the interior of the Sun. The large stellar distance, the point source character of the stars, the low-amplitude of the oscillations and effect of the earth's atmosphere on the signal restrict the asteroseismic studies to the use of the small sets of data often characterized by modes with only low-harmonic degree ($l<$3).  As we know the Sun oscillates in thousands of non-radial eigenmodes, and this richness in the number of identified modes leads to the great success of helioseismology using which one can probe the invisible internal structure of the Sun in detail. Using the ground based Birmingham Solar-Oscillations Network (BiSON, Chaplin et al. 1996) more than 70 independent modes with amplitude of few ppm and periods between 3--15 minutes were detected and identified as low-degree ($l$) pulsation modes ($l\leq$ 3) that can be used to probe the solar core (Chaplin et al. 2007). Detection of solar-like oscillations in stars is, however, very difficult because of their small amplitudes. 
The study of stellar oscillations (asteroseismology) have several advantages over the other observables because frequencies, amplitude and phase of the oscillations can be measured very accurately which depends upon the equilibrium structure of the model. Different modes of pulsations are confined and probe different layers of the interiors of the stars, thus accurate measurements of acoustic frequencies can be used not only to study the stellar interiors but also constraints on the theories of stellar evolution.

The measurements of basic physical parameters such as luminosity, effective temperature, surface composition and $v sin i$ (from spectrum analysis) age (estimated through stars in cluster), masses and radii (measured using the spectroscopic binaries) are affected by unknown effects such as loss, accretion and diffusion of mass. Therefore, the structure of the model of the stars can not be so well constrained such as that of the Sun. As a result, the precision attained by asteroseismology is much less than that obtained by helioseismic studies.  Nevertheless, several attempts have been made during recent times with the aim to identify oscillations in distance stars and also to model the stellar pulsation phenomena.  In particular, photometric observations obtained from space by the 
$ MOST\footnote{http://www.astro.ubc.ca/MOST/}, CoRoT\footnote{http://corot.oamp.fr/}$ and $~Kepler\footnote{http://www.kepler.arc.nasa.gov/}$ missions have led to in-depth understanding of asteroseismology by detection of  hundreds of stars with solar-like oscillations. 

\section{Asteroseismology}

Just as helioseismology has led to new insights on the internal structure
of the Sun, asteroseismology provides wealth of information on the
physical properties of many different types of stars.  To extract this
information, the observed pulsational frequencies are compared to the
frequencies predicted by a stellar model.  The input physical parameters 
of the models are adjusted to agree with observations.  If a model succeeds 
in reproducing the observed frequencies, it is called a seismic model. 
The applications of asteroseismology are used: (1) for the accurate and precise estimates of the stellar properties (i.e., density, surface gravity, mass, radius and age); (2) an additional information on the internal rotation and angle of inclination can also derived for a better understanding of the dynamics of the star; (3) to estimate the angle between the stellar
rotation axis, the line-of-sight, magnetic axis and pulsation axis, (4) to bring additional constraints on stellar models which allow more precise estimation of age, and (5) to derive the strength of magnetic field.

\section{Stellar Pulsations}

The pulsation of a star is characterized by expansion followed by contraction. If 
the pulsating star preserves its spherical symmetry then  mode of pulsation is
known as the radial. Otherwise the pulsation called as non-radial.  Many stars, including the Sun, pulsate simultaneously in many 
different radial and non-radial modes. Stars such as Cepheids and RR Lyra 
pulsate either in one or two radial modes. The physical basis for understanding stellar pulsations has been described with great clarity and a great detail by many authors (see, for instance, Cox 1980; Unno et al. 1989; Christensen-Dalsgaard \& Berthomieu 1991; Brown \& Gilliland 1994; Aerts et al. 2010; Balona 2010). Here we present only the
minimal description necessary for our purpose.

When a small contraction occurs, the local temperature slightly increases, inducing an enhancing the outward pressure. Hence, the layer expands again until the equilibrium is attained. The time requires to recover from such a small contraction is given by the
dynamical time-scale of the star:

\begin{equation}
\label{t_dyn}
t_{\rm{dyn}}=\left(\frac{R^3}{GM}\right)^{1/2}\propto\left(G\,\overline{\rho}\right)^{-1/2} \, 
\end{equation}

where $R$, $G$, $M$ and $\overline{\rho}$ are the radius, Gravitational constant, 
mass and mean stellar density, respectively. The periods of the oscillations 
generally scale to $t_{\rm{dyn}}$ which is expressed as the time star needs to go 
back to it's hydrostatic equilibrium if some dynamical process disrupts the balance 
between pressure and gravitational force. For a star close to hydrostatic 
equilibrium, this time scale is equivalent to the time it takes a sound wave to 
travel from the stellar center to the surface which is equivalent to the free-fall time-scale 
of the star). The period of oscillation of a star can not exceed the dynamical time. 
For the Sun, $\tau_{dyn} \sim$ 20-min while for a white dwarf this value is less 
than a few tens of seconds. Thus estimation  of the period immediately gives the important parameters of the star, viz. the mean density.

\begin{figure}
\includegraphics[width=1.1\textwidth,height=0.9\textheight]{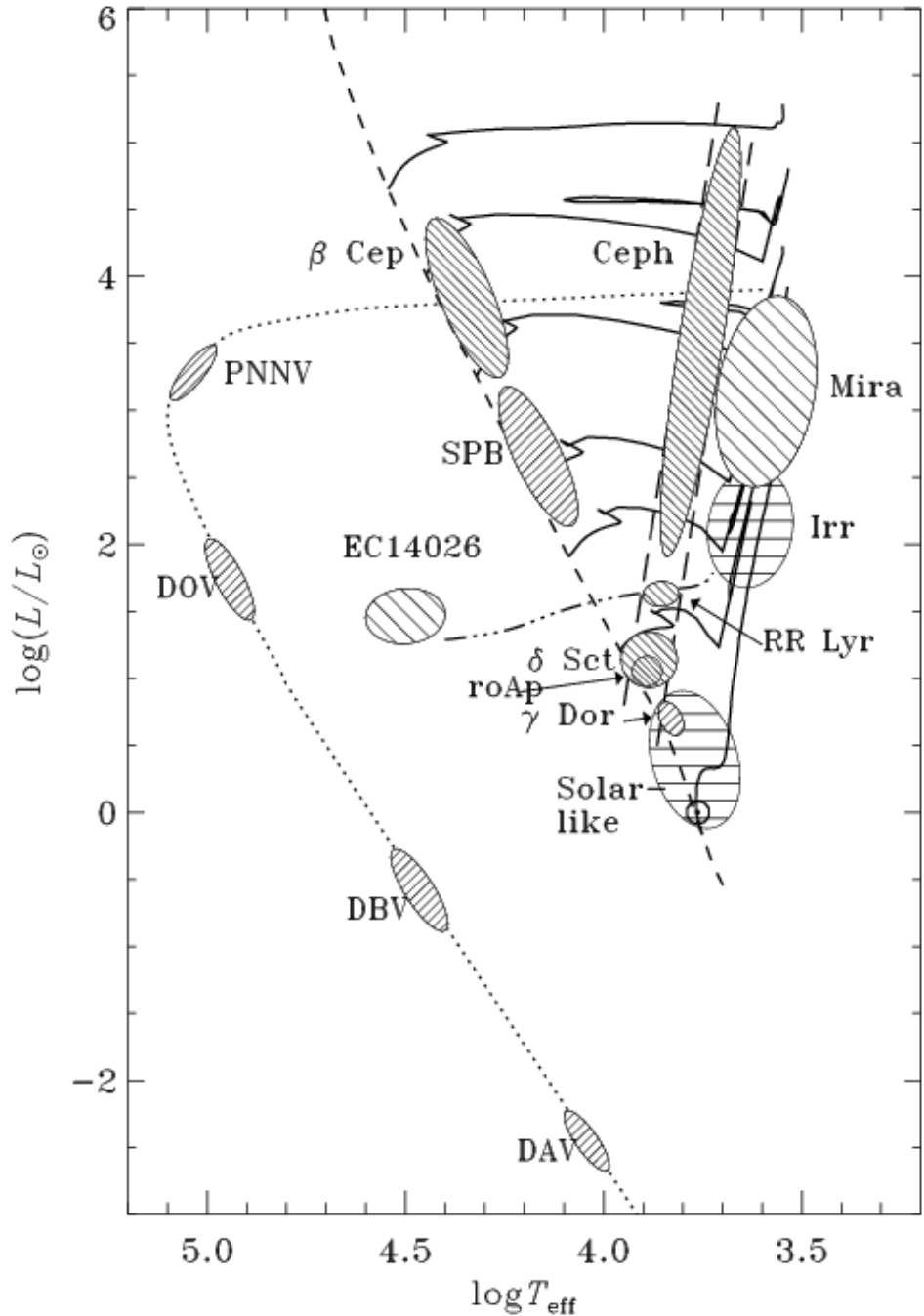}
\caption{: Theoretical HR diagram schematically illustrating  locations of known
pulsating stars. The dashed line marks the zero-age main sequence (ZAMS) 
where the solar-like oscillators, $\gamma $ Doradus, $\delta$ Scuti and roAp 
stars are located. The dotted line corresponds to the cooling sequence of 
white dwarfs, where one finds active planetary-nebula nuclei variable, variable white dwarfs. The parallel long-dashed 
lines indicate the Cepheid instability strip where RR Lyrae and Cepheids are 
situated. Adapted from Christensen-Dalsgaard (2000).}.
\label{HR}
\end{figure}

Fig. \ref{HR} shows a theoretical pulsational Hertzsprung-Russell (HR) diagram. 
Beside main-sequence (MS) pulsators (hydrogen burning in the core), there are  
pre-main sequence stars (nuclear reactions have not started yet) and 
post-main sequence stars (hydrogen burning takes place in a shell). 
The MS pulsating variables are excellent asteroseismic tools because 
most of them are multi-periodic. Along the MS one finds the early B and late O stars (M $\sim$ 8--20 M$_\odot$) there are the
$\beta$ Cephei pulsators with periods of hours. The long-period slowly pulsating B (SPB)  stars  
(M $\sim$ 3--12 M$_\odot$) are mid- to late B stars with periods of days. 
Moving towards lower masses, there are the $\delta$ Scuti pulsators 
(M $\sim$ 1.5--2.5 M$_\odot$) which are dwarfs or giants of spectral type 
A2--F5 located in the extension of the Cepheid instability strip and pulsate with  period of 0.02--0.3 d. Chemically peculiar (CP) pulsating magnetic stars of spectral type A are known as rapidly oscillating Ap (roAp) pulsators with period between $\sim$ 5--23 min. Among the F-type stars there are the $\gamma$~Dor pulsators (M $\sim$ 1.4--1.6 M$_\odot$) that pulsate with  periods 0.3--3\,d. At ARIES we have been studying the MS pulsating stars 
asteroseismically and participate in whole earth telescope (WET) 
campaigns organized for the un-interrupted time-series observations of pulsating variables.

\section{Basic Properties of Stellar Oscillations}

Generally stellar oscillations are studied in the linear regime where one considers small displacements.  Owing to limitations on existing computers, non-linear regime can only be studied for radial pulsations where the equations are greatly simplified.  As a result we can  predict the oscillation frequencies for any given stellar model but not 
the amplitude of the pulsations.

The small amplitude, linear oscillations of a spherically symmetric
non-rotating star can be expressed in terms  of a spherical harmonic 
$Y_l^m(\theta,\phi)$ where  $\theta$ is co-latitude and $\phi$ is the longitude. 
The radial component of displacement can be expressed as :

\begin{equation}
\label{displacement}
\xi_r(r,\theta,\phi;t)=\Re\left\{a(r) \, Y_l^m(\theta,\phi) \, \exp(-{\rm{i}}\,2
\pi \nu t)\right\} \, ,
\end{equation}

where $r$ is the distance to the center of the star, $a(r)$ is an amplitude 
function and $\nu$ is the cyclic frequency of the oscillation.   For a 
spherically symmetric star the frequency of oscillations depend
on $n$ and $l$, i.e., $\nu\!=\!\nu_{nl}$. 

The spherical harmonic $Y_l^m(\theta,\phi)$ is expressed as :
\begin{equation}
\label{spharm}
Y_l^m(\theta,\phi)=(-1)^m \, c_{lm} \, P_l^m(\cos \theta) \, \exp({\rm{i}}\,m\phi) \, ,
\end{equation}
where $P_l^m$ is an associated Legendre function given by
\begin{equation}
P_l^m(\cos \theta)=\frac{1}{2^l l!} \, (1-\cos^2 \theta)^{m/2} \, \frac{{\rm{d}}
^{l+m}}{{\rm{d}}\cos^{l+m} \theta} (\cos^2 \theta -1)^l \, ,
\end{equation}
and the normalization constant $c_{lm}$ is determined by
\begin{equation}
c_{lm}^2=\frac{(2l+1)(l-m)!}{4\pi (l+m)!} \, ,
\end{equation}
such that the integral of $|Y_l^m|^2$ over the unit sphere is unity.

\begin{figure}
\center{
\includegraphics[width=0.6\textwidth]{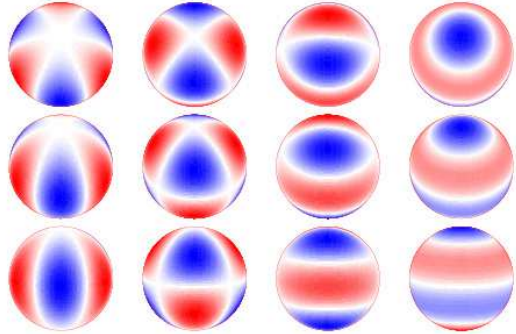}
}
\caption{: Snapshot of the radial component of the $l$ = 3 octupole modes. 
Each row display the same modes with different inclination angles 
of the polar axis with respect to the line of sight: $30^\circ$ (top row), 
$60^\circ$ (middle row), and $90^\circ$ (bottom row). White bands represent 
the nodal surface lines, red and blue sections represent portions of the stellar 
surface that are moving in and out, respectively. The right-most column displays 
the axisymmetric (i.e., with $m\!=\!0$) mode $(l\!=\!3,m\!=\!0)$. From right to 
left, the middle columns display the tesseral (i.e., with $0\!<\!|m|\!<\!l$) 
modes $(l\!=\!3,m\!=\!\pm1)$ and $(l\!=\!3,m\!=\!\pm2)$. The left-most column 
displays the sectoral (i.e., with $|m|\!=\!l$) mode $(l\!=\!3,m\!=\!\pm3)$. 
Adopted from Kurtz et al. (2006).}
\label{spher_harm}
\end{figure}

Each eigenmode of a  spherically symmetric star is specified by the three
indices, $n, l$ and $m$, which are called  the radial order, the spherical
degree and the azimuthal order, respectively. The radial order $n$ is associated
with the structure in the radial direction, the indices $l$ represents the number 
of the surface nodes or lines in the direction of equator and $m$ defines the 
number of surface nodes that are lines of longitude (Takata 2012). The modes with $l$=0, 1 and 2 correspond to radial, dipolar and quadrupole-polar modes, respectively. The modes with $m\!\neq\!0$ are traveling waves.  Modes traveling in the direction of rotation are 
called prograde modes (positive $m$) while waves traveling 
in the opposite direction are called retrograde modes (negative $m$). Low-overtone 
modes have $n<5$ and high-overtone modes have $n > 20 $. Fig.\, \ref{spher_harm} illustrates the appearance of the $l\!=\!3$ octupole modes on a stellar surface.

There are $2l+1$ values of $m$ for each value of degree $l$. In the case of a 
spherically symmetric, non-rotating star their frequencies will be the same. 
Rotation, as well as any other physical process results in a departure 
from spherical symmetry that introduces a dependence of the frequencies of 
non-radial modes on $m$. When the cyclic rotational frequency of the star, 
$\nu_{rot}$ is small and in the case of rigid-body rotation, the cyclic 
frequency of a non-radial mode is given to first order (Ledoux 1951):

\begin{equation}
\nu_{nlm} = \nu_{nl0} + m \nu_{rot} , \vert m \vert \leq l 
\end{equation}

High-degree modes penetrate only to a shallow depth while low-degree 
modes penetrate more deeply. If many modes are present then a range of
depths can be investigated.  It is then possible to ``invert'' the observations to 
make a map of the sound speed throughout the star.  One may thus deduce the 
temperature and density profile using reasonable assumptions about the 
chemical composition.

In a non-radial mode, some parts of the stellar surface are brighter and
others fainter.  Some parts are moving towards the observer and others away
from the observer.  As a result, the nullifying effect reduces the observed
light and radial velocity amplitude.  The higher the value of $l$, the
greater the nullifying effect.  For most stars observed from the ground, only
modes with $l < 4$ are typically detectable.

\subsection{Pressure and Gravity Modes}

There are two main sets of solutions of the equations of motion for a pulsating 
star leading to two types of pulsation modes namely $p$- and $g$-modes. For the 
$p$- or pressure modes, pressure acts as the restoring force for a star perturbed 
from equilibrium. The  $p$-modes are acoustic waves and have motions that 
are primarily vertical. For the $g$- or gravity modes buoyancy acts as a restoring 
force and gas motions are primarily horizontal. 

There are two other important properties of $p$-modes and $g$-modes: (1) as the 
radial overtone ($n$) increases, the frequencies of the $p$ modes increase, but the 
frequencies of the $g$-modes decrease; (2) the $p$-modes are sensitive to 
conditions in the outer part of the star, whereas $g$-modes are sensitive to 
the core conditions. Since buoyancy demands motions that are primarily horizontal, 
there are no radial (i.e., $l\!=\!0$) $g$-modes. As stars evolve, the convective 
envelope expands and the acoustic oscillation modes ($p$-modes) decrease in 
frequency. At the same time, $g$-mode oscillations that exist in the core of the 
star increase in frequency as the core becomes more centrally condensed. Eventually, 
$p$- and $g$-mode frequencies overlap, resulting in oscillation modes that
have a mixed character, behaving like $g$-modes in the core and $p$-modes in the 
envelope.

To identify the $p$-mode radial overtones, it is useful to introduce a quantity 
known as the pulsation constant $Q$ defined as :

\begin{equation}
Q = {P_{\rm osc}}\sqrt{\frac{\overline{\rho}}{\overline{\rho}{_{\odot}}}}
\label{eq:1}
\end{equation}
where $P_{\rm osc}$ is the pulsation period and $\overline\rho$ is the mean density 
of the star.

Equation\,(\ref{eq:1}) can be rewritten as

\begin{equation}
\log Q = -6.454 +\log P_{\rm osc} +\frac{1}{2}\log g +\frac{1}{10}M_{\rm bol} +
\log T_{\rm eff},
\label{eq:2}
\end{equation}

where the unit of $P_{\rm osc}$, $\log g$ and $T_{\rm  eff}$ are iin unit of days, dex and 
Kelvin, respectively. Table \ref{Q} lists the value of $Q$ for a
$\delta$~Scuti star pulsating in different modes. The value 
of $Q$ is not too sensitive to $M_{\rm bol}$ or $T_{\rm eff}$, but it is sensitive 
to $\log g$, which is the most  uncertain quantity. For standard late-A and early-F 
star models, we expect $Q$=0.033 and $Q$=0.025 for the fundamental and first-overtone mode (Stellingwerf 1979), respectively. Breger \& Bregman (1975) demonstrated 
that for $T_{\rm eff}$ $<$ 7800 K, the highest amplitude mode tends to be the fundamental 
mode, whereas for hotter temperatures the first overtone is prevalent.

\subsubsection{The Asymptotic Relation}

If the condition $l\ll$ $n$ is satisfied, the excited $p$-modes in a
non-rotating star can be described according to the asymptotic theory 
(Tassoul 1980), which predicts that oscillation frequencies 
$\nu_{n,l}$ of acoustic modes, characterized by radial order ($n$) and 
harmonic degree ($l$) satisfy the following approximation:

\begin{equation}
\label{asympt}
\nu_{nl} \approx \Delta\nu(n + \frac{1}{2}l + \epsilon) - \delta\nu_{n,l}
\end{equation}

where $\epsilon$ is a constant sensitive to the surface layers.  From observational point of view it is convenient to represent the frequency spectrum by the average quantities $\Delta\nu = \nu_{n,l} - \nu_{n-1,l}$ and $\delta\nu_{n,l}=\nu_{n+1\,l}-\nu_{n,l}$ 

\begin{figure}
\caption{: Close-up of the $p$-mode amplitude spectrum of the Sun.
 The peaks are marked with the corresponding ($n,l$) values, which were determined 
by comparison with theoretical models. Adopted from Bedding \& Kjeldsen (2003).}
\center{
\includegraphics[width=1.1\textwidth,height=0.4\textheight]{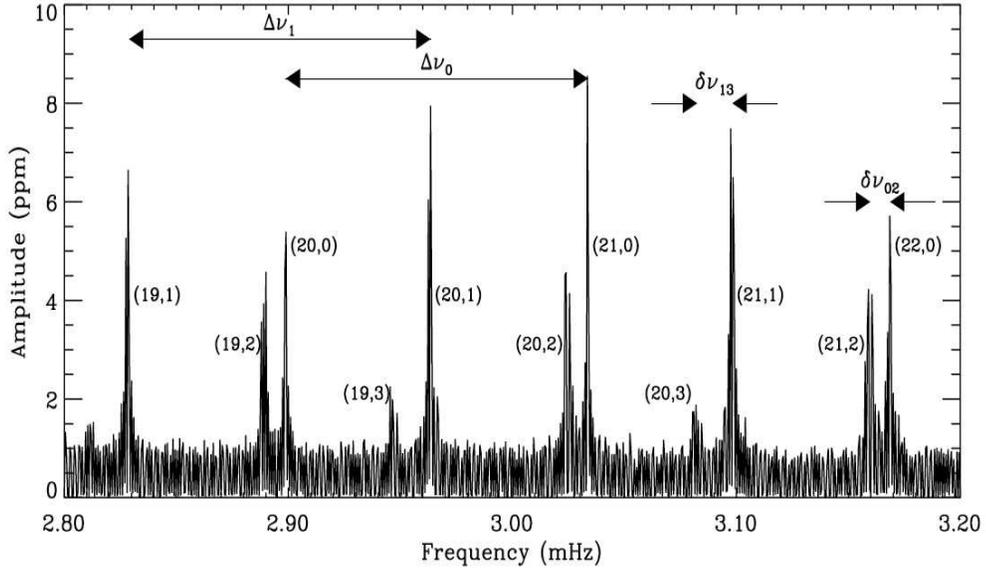}
}
\label{solarspec}
\end{figure}

The parameter $\Delta\nu$ which is the frequency difference between successive modes of the same $l$ is called the {\it large separation}. This is proportional to the inverse of the sound travel time across the stellar diameter. This is related to the sound speed profile in the star and proportional to the square root of the mean density of the star :

\begin{equation}
\Delta\nu_0 = \left(2\int_0^R\frac{{\rm{d}}r}{c}\right)^{-1} \propto \sqrt{\overline{\rho}}
\end{equation}

where  $c$ is sound speed, $\overline{\rho}$ is mean density and $R$ is the stellar radius.
The parameter $\delta\nu_{n,l}$ is known as the {\it small separation} and is sensitive to the core region of the star which in turn is sensitive to the composition profile. Thus the small frequency separation is an important diagnostic of the stellar evolution.    
The large and the small separations  in the acoustic amplitude spectrum of the Sun is shown in Fig.~\ref{solarspec}.

\begin{figure}
\center{
\includegraphics[width=1.1\textwidth,height=0.55\textheight]{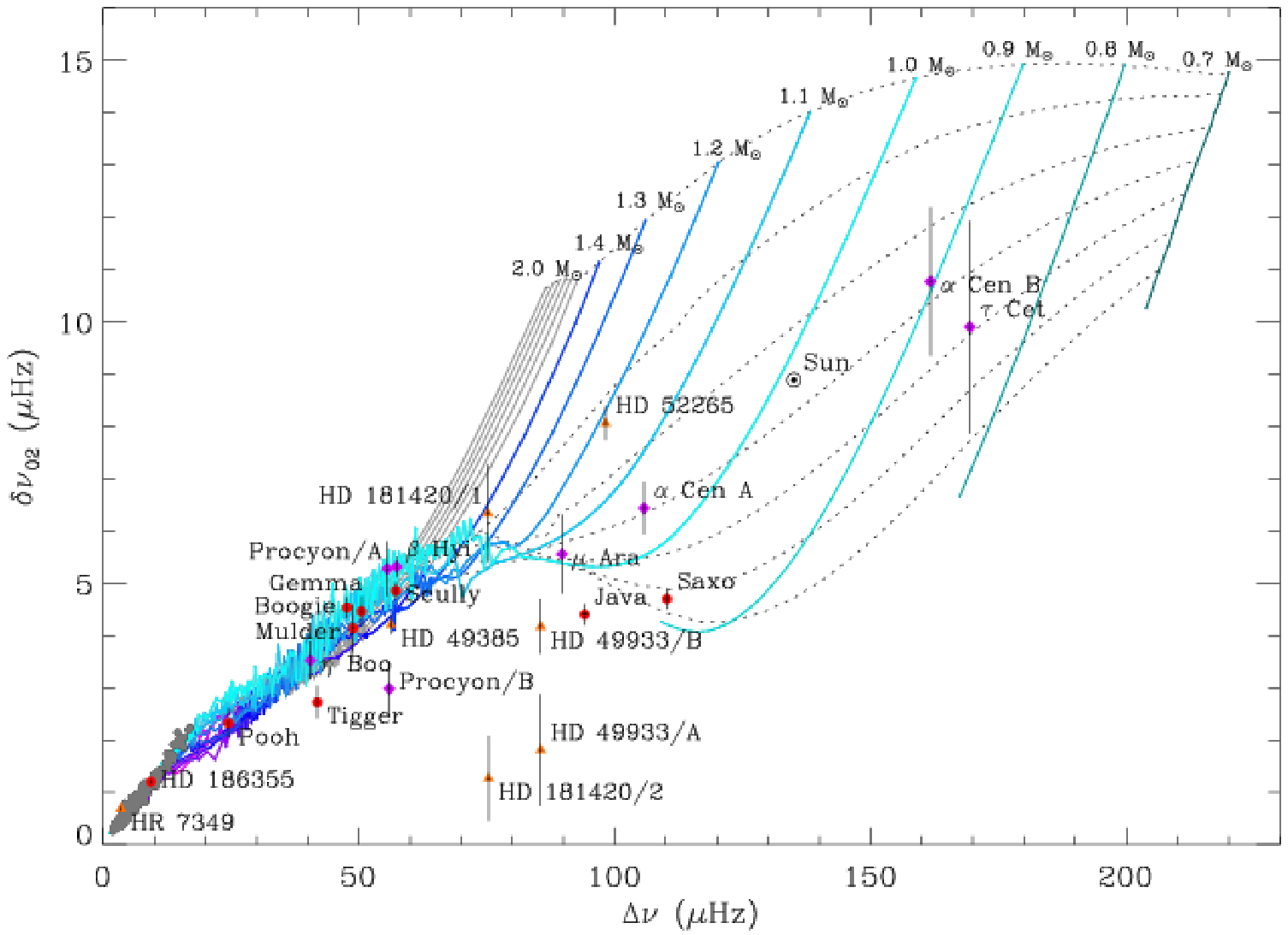}
}
\caption{: C-D digram showing the large separation versus the small separation 
for a series of evolution models calculated for different masses with model tracks for near-solar metallicity ($Z_0$ = 0.017).  The section of the evolutionary tracks in which the models have a higher $T_{eff}$ than the approximate cool edge of the classical instability strip (Saio \& Gautschy 1998) are gray. Dashed black lines are isochrones, increasing by 2 Gyr from 0 Gyr (ZAMS) at the top to 12 Gyr at the bottom. Stars observed by CoRoT (orange triangles), Kepler (red circles) and from the ground (purple diamonds) are marked. Gray circles are Kepler red giants (Huber et al. 2010) and Sun is marked by its usual symbol.  Adopted from White et al. (2011).}
\label{CD}
\end{figure}

Measuring the large and small separations gives a measure of the mean density
and evolutionary state of the star. A useful seismic diagnostic is the 
asteroseismic HR diagram where the average large separation is plotted 
against its average small separation and known as the ``C-D'' diagram (Christensen-Dalsgaard 1984). Fig. \ref{CD} shows the C-D diagram for models with near-solar metallicity. The solid lines show the evolution of $\Delta \nu$ and $\delta \nu_{02}$ for models with a metallicity close to solar and various masses. Isochrones of different age are also shown with dashed lines. For MS, the asteroseismic C-D diagram allows us to estimate the mass and age of the star  assuming that other physical inputs such as initial chemical composition and the convective 
mixing-length parameter are known. As stars evolve off the MS, their tracks converge for the sub-giant and red-giant evolutionary stages.

The high-order $g$-modes are observed in $\gamma$-Dor and SPB stars 
as well as in white dwarfs.  In the absence of rotation, the periods 
$\Pi_{n,l}$=1/$\nu_{n,l}$ are approximately uniformly spaced and to first
order satisfy the relation :

\begin{equation}
\Pi_{n,l} = \frac{\Pi_0(n+\frac{l}{2}+\alpha_g)}{\sqrt{l(l+1)}}
\label{asyg}
\end{equation}
where  $\Pi_0 = 2 \pi ^2 \left (\int_{r_1}^{r_2} \frac {N}{r} dr \right)^{-1}$ 
is the basic period spacing, $N$ is the buoyancy (or Brunt-Vaisala) frequency 
and [r1, r2] is the interval where the modes are trapped.  $\alpha$ is a phase 
constant that depends on the details of the boundaries of the $g$-mode trapping 
region and the integral is computed over that same region. Departures from the 
simple asymptotic relation given in Eq. \ref{asyg} is used as a means to 
diagnose the stratification inside stars such as white dwarfs.  

\subsection{Identification of Pulsation Modes}

Successful seismic modeling involves not only detection of pulsation frequencies 
but also relies on the identification of the spherical harmonic numbers ($n,l,m$) 
of the observed modes. Two particularly promising methods for identifying pulsation modes are multi-colour photometric and spectroscopic mode identification. The first approach consists of measuring the amplitudes and phases of a given pulsation mode in different photometric
bands, calculating the ratios of the different amplitudes and/or the phase differences, and comparing to theoretical predictions. The value of $l$ can be determined from the amplitude ratios and phase differences of the variation at different wavelength bands (Garrido et al. 1990). The second approach exploits the Doppler shifts caused by the velocity field from the pulsation mode and how it affects observed
absorption lines. These methods have been successfully applied to slowly rotating stars (e.g. De Ridder et al. 2004; Zima et al. 2006; Briquet et al. 2007), but more work is required  before they are applied to rapid rotators.  High-resolution spectroscopy can be used to study 
the line profile variations (LPVs) during the pulsation.  The observed
profile variations are compared with the profile variations modeled for
different values of $l$ and $m$.

\begin{figure}
\center{
\includegraphics[width=1.0\textwidth,height=0.20\textheight]{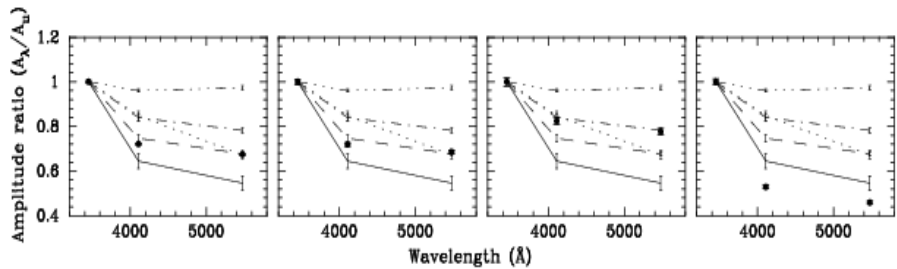}
}
\caption{: Identification of the four strongest pulsation modes of the
$\beta$~Cephei star  12~Lacertae using multi-colour photometry. Adopted from 
Handler et al. (2006).}
\label{mode}
\end{figure}

\begin{figure}
\center{
\includegraphics[width=1.0\textwidth,height=0.30\textheight]{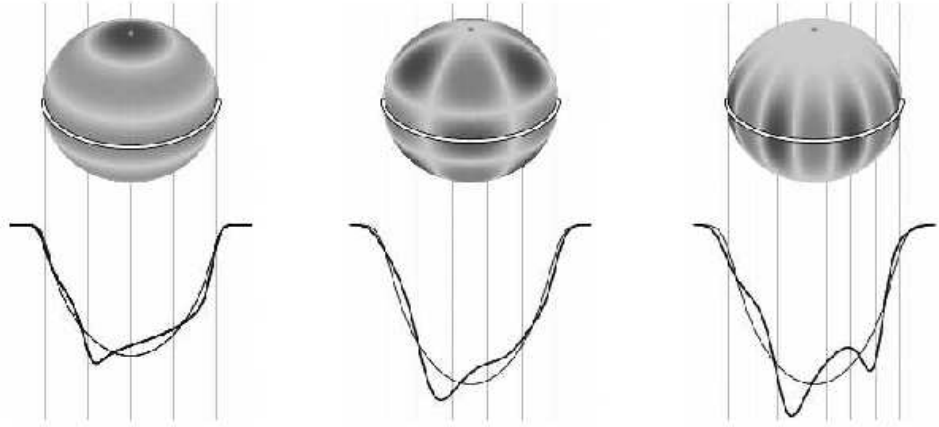}
}
\caption{: Line profile variations due to stellar pulsation. The upper parts of the graph
shows the shape of the oscillation mode on the surface, whereas the thin lines in the
lower halves represent the unperturbed rotationally broadened line profile, and the thick
lines are the superpositions with the pulsation. Each individual mode (from left to right:
$l$ = 4,$m$ = 0; $l$ = 5, $|m|$ = 3; $l$ = $|m|$ = 7) generates a different distortion 
of the line profile. Adopted from Telting \& Schrijvers (1997).}
\label{doppler}
\end{figure}

Photometric mode identification is widely used  because it is very effective 
and easy to use if the observations are sufficiently precise. The amplitude 
ratios can be calculated following the method of Dupret et al. (2003) and an
example of such a mode identification is shown in Fig. \ref{mode}. 
The spectroscopic method of mode identification relies on the Doppler
effect.  The intrinsic line profile is blue-shifted in the parts of the 
stellar surface approaching the observer and red-shifted on those parts
moving away from the observer. The effect is strongest on the stellar limb 
and decreases towards the center.  Hence,  a rotationally broadened line 
profile contains spatial information of the brightness distribution on
the stellar surface such as that caused by stellar pulsation.  The method 
of reconstructing the  stellar surface brightness distribution
from line profile variations is called Doppler Imaging (Telting 2003). By examining a stellar line profile pulsation modes up to 
$l \approx $20 can be observed and identified. One example of pulsational 
line profile variations is shown in Fig. \ref{doppler}.

High-resolution spectroscopy allows us to study LPVs using different methods.
To study the LPVs from the spectroscopic data one should carefully select a 
number of unblended, deep lines and avoid saturated lines. Averaging all 
lines in the spectrum (Uytterhoeven et al. 2008) provides high S/N mean line 
profile.  This can be done using cross-correlation. The drawback of the
spectroscopic method is that it assumes that all the lines have the same 
intrinsic profile and that they form in the same region of the atmosphere, 
which may not be the case.  In spite of the fact that high-dispersion 
spectroscopy potentially contains  more information than multi-colour 
photometry, most mode identifications are performed  from multi-colour
photometry.  Spectroscopy is mostly used for the confirmation of the mode identified using photometry (Daszy{\'n}ska-Daszkiewicz et al. 2005).

\begin{figure}
\center{
\includegraphics[width=1.0\textwidth,height=0.30\textheight]{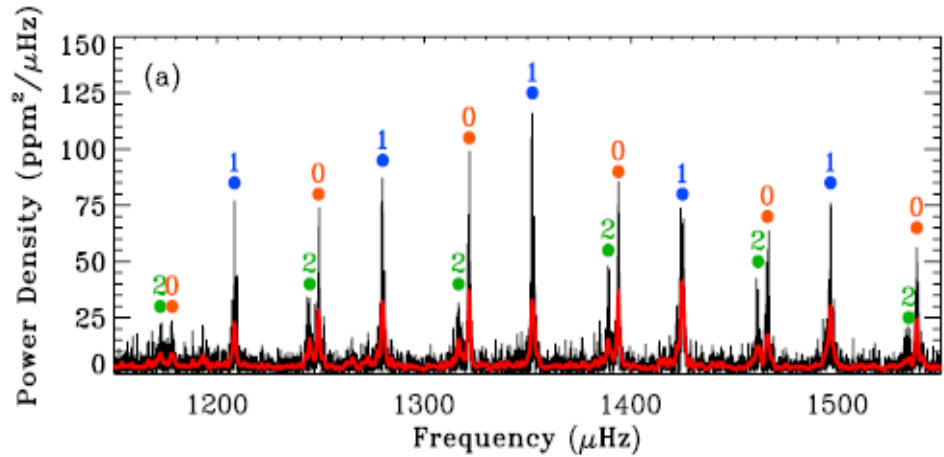}
}
\caption{: Power spectra of  a G-type star KIC 6933899. The red curves show the power spectra after smoothing. Mode identification of the G star is trivial, with modes of $l$ = 0 (orange), 1 (blue), and 2 (green) labeled. Adopted from White et al. (2012).}
\label{misol}
\end{figure}

 More recently Ligni{\`e}res \& Georgeot (2009) calculated geometrical disk integration factors of acoustic modes in deformed polytropic models by integrating the temperature fluctuations over the visible disk. The effects of rotation were fully taken into account in the pulsation modes, thanks to the 2-D numerical approach, but non-adiabatic effects were neglected, thereby making the fluctuations of the effective temperature inaccessible. 

In the Sun and solar-like oscillators the mode identification is straightforward
thanks to the distinctive pattern of alternating odd and even
modes in the power spectrum. Each $l$ = 0 mode is separated by $\delta \nu_{02}$ from an $l$ = 2, and separated by $\Delta \nu$ /2 - $\delta \nu_{01}$ from the $l$ = 1 mode of the same order. An example is shown in Fig. 7 for the Kepler star KIC 6933899 of 
effective temperature of 5840 K similar to the Sun (White et al. 2012). Fig. 3 shows the identification of the modes from the solar spectrum. 

\begin{figure}
\center{
\includegraphics[width=0.8\textwidth,height=0.40\textheight]{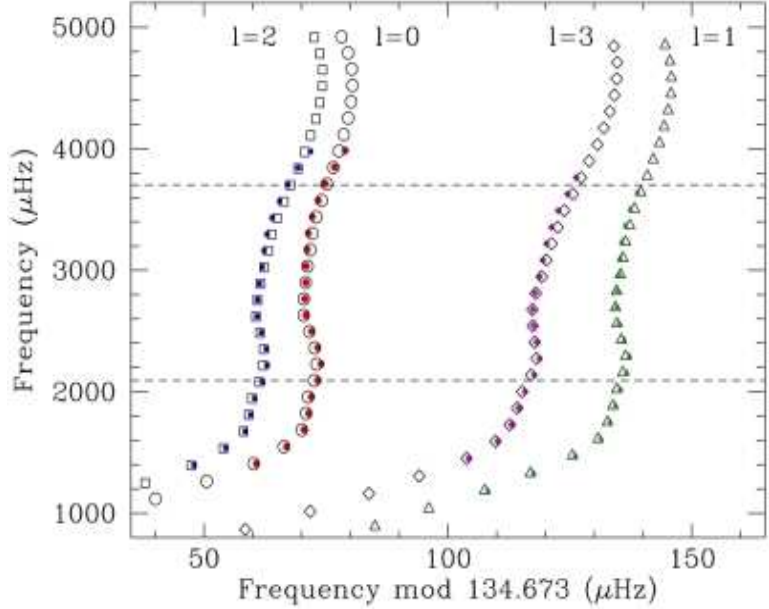}
}
\caption{: An {\`e}chelle diagram for the Sun observed as a star, where we divide the oscillation spectrum into segments of a fixed length and plot them against the oscillation frequency. Colored symbols show the observations while open points show the best stellar model. From left to right the ridges are paired $l$ = 2 then $l$ = 0. Followed by the odd pair of $l$ = 3 and $l$ = 0. Adopted from Metcalf et al. (2009).}
\label{ech-sol}
\end{figure}

The {\`e}chelle diagram can be used to identify the angular degree of an observed frequency.
This process was first suggested by Grec et al. (1983) and used to identify the degree of
solar modes. The process involves splitting a spectrum into lengths of the star’s large
separation which are then stacked on top of each other. This process has the effect of
lining up frequencies into ridges corresponding to different angular degrees. Fig. 8 shows an example of an {\`e}chelle plot created from frequencies found in data collected by
BiSON network. In this figure we can see that the frequencies are splitted into four distinct regions.

\section {Pulsational Mechanism} 

The mechanism which drives stellar oscillations can be inferred by studying
and modeling the energy transport in the interior of a star.  We have
mentioned three types of oscillating star: stars exhibiting solar-like 
oscillations (including the Sun itself), Cepheid variables and Mira. 
Pulsations in these three different types of stars are driven by different
excitation mechanisms.  

Self-driven oscillations, such as those in Cepheid variables, can be either 
intrinsically unstable or intrinsically stable depending upon the
mechanism.  In the former case, oscillations result from the amplification of 
small disturbances by means of a heat-engine or valve mechanism converting thermal
energy into mechanical energy in a specific region usually a radial layer of 
the star (Zhevakin 1963). During the compression phase heat is absorbed, while in the
expansion phase heat is released. In order to cause overall  excitation of 
the oscillations, the region associated with the driving has to be at an appropriate 
depth inside the star, thus providing an explanation for the  specific location of the 
resulting instability belt in the H-R diagram (Fig. \ref{HR}). Such a region 
is typically associated with a region of ionization of an abundant elements.
Normally, the opacity decreases as the temperature increases.  In most
regions of a star, compression  results in an increase in temperature.  The 
resulting  decrease in opacity contributes strongly to heat leakage and to the 
stability of the star as a whole.  However, in a zone where an abundant atomic species 
is partly ionized, the  opacity  increases with temperature because radiation is 
absorbed by ionization of the material. Upon compression, some of the heat is 
absorbed in ionizing more of the atomic species. Thus energy is absorbed on 
compression and released on expansion. The cycle repeats and this process is known as the $\kappa$-mechanism. It is believe that $\kappa$-mechanism is responsible for the pulsations observed in stars that are located on the instability strip: classical Cepheids, WW Virginis, RR Lyrae and $\delta$ Scuti stars. The pulsation mechanism for other stars outside the instability strip such as  Miras, semi-regular variables is not so well understood.

Solar-like oscillations are expected in cool main-sequence, sub-giant and red-giant stars ($0.1 M_{\odot} < M < 8 M_{\odot} $ (Christensen-Dalsgaard \& Frandsen 1983; Houdek et al. 1999; Chaplin et al. 2011a,b; Hekker et al. 2011; Chaplin \& Miglio 2013)). It is popularly accepted that the five-minutes oscillations of the Sun are excited by the stochastic
effect of turbulence (Goldreich \& Keeley 1977;
Kumar \& Goldreich 1989; Goldreich et al. 1994; Belkacem et al. 2008). As a natural extension,
solar-like oscillations of stars have also been considered to be due to the same stochastic excitation of turbulence (Samadi \& Goupil 2001; Samadi et al. 2003; Samadi et al. 2008).

\section{Measurements of Variability}

Stellar oscillations generate motions and temperature variations causes light, 
radial velocity and line profile changes. Pulsating stars can be studied both 
photometrically and spectroscopically, via time-series measurements. 
The angular diameters of stars are too small to resolve surface features
and we must rely on integrated light or radial velocity variations of the
visible hemisphere of the star.  In the following sub-sections we briefly describe only two 
methods of obtaining asteroseismic information.

\subsection{Photometric Technique}

The integrated stellar intensity can be measured using photoelectric and CCD photometry.  The
field of view allows the simultaneous observation of many stars.  The major source 
of atmospheric noise in photometric observations is sky transparency variations. On a 
photometric night the sky transparency variations occur on times-scale of about 
15-min and longer giving rise to low-frequency noise. This noise can be be partially 
controlled by observation of non-variable comparison stars or observing a star under  photometric sky condition. The second major source of atmospheric noise in photometry is scintillation caused by variable refraction. In practice, scintillation noise drops
inversely with telescope aperture  (Dravins et al. 1998). Fig. 9 shows the variation of the scintillation noise with the diameter of telescope which clearly reveals that the scintillation noise can be reduced by observing with a bigger telescope. 

\begin{figure}
\center{
\includegraphics[width=0.6\textwidth,angle=90]{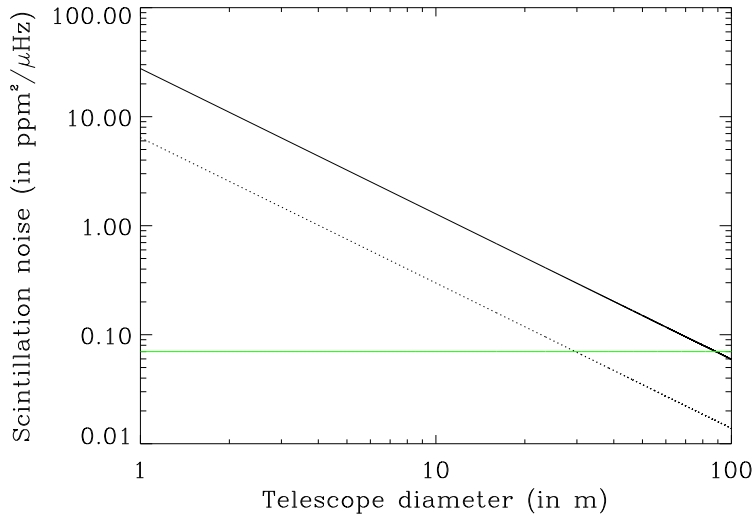}
}
\caption{: Scintillation  noise as a function of telescope diameter for an air-mass of $X=1.5$ (solid line)  and of $X=1.0$ (dotted line) for an observatory located at altitude of 4000 m. The green line is the limit of the scintillation noise for  the solar-like oscillations. 
Adopted from Appourchaux \& Grundahl (2013).}
\label{scinti}
\end{figure}

\subsection{Spectroscopic Technique}

The measurement of radial velocity (RV) from a single absorption line is a challenging 
task because of the lack of photons. Therefore, the spectrograph
should cover a wide wavelength range at high-resolution with minimum optical loss.
The spectrographs such as High Accuracy Radial Velocity Planet Searcher (HARPS) achieve high-spectral stability by controlling of the 
air pressure and temperature in the instrument (Pepe et al. 2000).
The precision reached by such an instrument depends primarily on  the number of photons gathered (telescope diameter, optical efficiency); the spectral coverage and the resolving power.

An added benefit of RV compared to light intensity measurements is
that the canceling effect for non-radial oscillations is smaller, allowing
modes of high $l$ to be observed and to detect modes of life times up-to a year 
(Salabert et al. 2009).  A combination of spectroscopic and photometric
observations, preferably simultaneously is of great importance for correct
interpretation of the data.  Such studies allow to determine the phase lag between 
light and RV curves, an important parameter for modeling of stellar  structure.
Intense observing campaigns that combined ground based spectroscopy and space-based 
photometry obtained with the MOST satellite were organized for the asteroseismic study of roAp
stars HD\,24712 (Ryabchikova et al. 2007), 10~Aql (Sachkov et al. 2008) and 
33~Lib (Sachkov et al. 2011).

\section{Ground and Space based Observations}

Different types of data sets and techniques are used for asteroseismology.  A
few milestones that have occurred in the past few years  are: 1) Small and medium-sized telescopes all over the world have been 
collecting optical photometry with photoelectric photometers and CCDs detectors; 
2) co-ordinated multi-site campaigns such as WET set-up by different 
communities have been used to obtain the continuous observations. These 
networks have improved the precision of the frequencies of pulsation by minimizing 
the aliasing; 3) in the past few years  the development of high-accuracy 
spectrographs such as HARPS and Ultraviolet and Visual Echelle Spectrograph (UVES) have allowed the detection and eventually 
confirm the presence of solar-like oscillations in stars other than Sun. 
These efforts have contributed to the refinement of global parameters of
stars and provided the mode identification; 4) large surveys such as Optical Gravitational Lensing Experiment ({\it OGLE}) and Massive Compact Halo Object ({\it MACHO}) have detected numerous pulsating variables; and  5) Interferometers 
such as Center for High Angular Resolution Astrometry ({\it CHARA}) and Very Large Telescope Interferometer ({\it VLTI}) have contributed significantly in asteroseismology  in terms of  
providing accurate radii for the pulsating variables. 

The precision obtained from ground-based photometric observations is not sufficient to
detect the $\mu$-mag amplitudes of the solar-like oscillations.
From space, much lower detection levels were first obtained on whole-disk
observations of the Sun (Woodard \& Hudson 1983). The space-based instruments such as {\it Wide Field Infrared Explorer (WIRE)} (Buzasi et al. 2000), $\it MOST $ (
Matthews et al. 2004) discovered many stars exhibiting the solar-like oscillations. Asteroseismic space missions have revolutionized the field asteroseismology with the launch of $\it CoRoT$ (Michel et al. 2008;  Auvergne et al. 2009; Baglin et al. 2009) and $\it Kepler$ (Chaplin et al. 2010, Koch et al. 2010). The primary aim of  both the missions were detection of exoplanets using the transit 
technique with secondary purpose to make the asteroseismic studies of pulsating variables.

The CoRoT satellite was launched on 27 December 2006 into an orbit around the Earth. This satellite had an off-axis telescope of diameter of 28~cm.  Four CCD detectors 
are mounted at the focal plane, covering  1.3 $^o$ $\times$ $1.3^o$ field of view for
each CCD.  Two of the CCDs are optimized for studying planet transits and the 
other two, with slightly defocused images, optimized for asteroseismology.
The detection of solar-like oscillations in hundreds of red giants was a 
landmark for the CoROT mission (De Ridder et al. 2009). 

The $\it Kepler$ mission was launched on 7 March 2009 into an Earth-trailing heliocentric orbit for a period of around 53 weeks and still operational. $\it Kepler$ observed a fixed field in the region of the constellations of Cygnus and Lyra, centered $13.5^o$ above the Galactic
plane. $\it Kepler$ has a Schmidt telescope with a corrector diameter  0.95-m. 
Array of 42 CCD array with a total of 95 megapixels covers  a field of around 
105 square degrees. The brightness variation on up to 170,000 targets
were available in the 29.4-minute long cadence (LC) almost continuously over four years.
A total of 512 main sequence and giants stars were observed with a 58.85-sec cadence over a time span of a few months. Chaplin et al. (2014) have presented a homogeneous asteroseismic analysis of more than 500 stars and derived their global asteroseismic parameters.

Temporal coverage  has a strong impact on detection of stellar variability. 
Gaps in the time-series data are obvious due to bad weather conditions and 
to day/night breaks.  Mosser \& Aristidi (2007) showed that since the local 
time approximately shifts by 4 min per day with respect to the sidereal time, the duration 
of an observation run on the same star cannot be longer than 5 months. The 
small temporal data due to interruptions generate daily aliases in the 
frequency spectrum located at 11.57 $ \mu$Hz and its harmonics.  These
aliases  make mode identification difficult when the frequency separation is about 
that value.  It is often suggested that the presence of such interruptions 
could be avoided by observing pulsating variables with a network of telescopes spread around the globe.  This applies to both stellar radial velocity and photometric measurements. The 
helioseismic ground-based networks that have been operating for last three decades are 
GONG (Harvey et al. 1996)) and BiSON (Chaplin et al. 1996)). Similarly the ground-based networks such as WET uses many telescopes all over the world.

The advantage of a space mission over ground-based observations is mainly in the
high precision and large sky coverage that can be achieved for long
durations without any interruption.  Another advantage of space observations 
is the possibility of using  wavebands which are blocked by the atmosphere 
(UV, EUV, X rays).  

The advantage of the ground-based spectroscopic observations is that it is possible to detect 
oscillation frequencies to much higher precision and at lower frequencies with 
radial-velocity observations.  Furthermore, it is possible 
to detect $l > 3$ modes through line profile  measurements. The ground-based photometric 
observations can be used for mode identification which require multi-band
observations currently not available from space.  The overwhelming advantage
of ground-based observations, of course, is that they are inexpensive.  A single 
space mission costs far more than a large telescope and can cover operating
expenses of ground-based observations for many years.

\section{Asteroseismic Techniques}

The determination of the pulsation frequencies is the first objective in any
asteroseismic investigation, accomplished by constructing the
periodogram which is the Fourier transform of time-series data that gives the power or amplitude as a function of time.  The
oscillation frequencies are seen as sharp peaks in the periodogram which can
be easily identified if the height of the peak is much larger than the
background noise level in the periodogram.  

For the frequency analysis of the classical pulsators traditionally, a process called ``pre-whitening'' is used to extract the frequencies from the periodogram and particularly it is useful to remove the unwanted frequency peaks which otherwise arise due to aliases or earth atmosphere.  This consists in identifying the frequency of a peak and removing a sinusoid with this frequency and to determine correct amplitude and phase from the data.  A periodogram of the pre-whitened data no longer shows a peak at this frequency. The process of successive
pre-whitening stops when the peak of highest-amplitude in the periodogram
has an amplitude which is considered too small to be significant.
A widespread criterion  used for this purpose is to calculate the ratio of
the peak amplitude to the background noise level.  When the S/N ratio
is less than 4, the signal is no longer considered statistically
significant (Breger et al. 1993). 

This technique works well for ground-based data where the S/N is quite
moderate.  For space observations of very high S/N, the prewhitening
technique introduces additional low-amplitude side-lobes due to inevitable
errors in frequency determination. For example the periodogram of solar-like oscillations 
exhibits the comb-like structure with amplitudes which decrease sharply from a central maximum and in such case to identify the location of the frequency of maximum
amplitude ($\nu_{max}$) is very important. A crude way of obtaining $\nu_{max}$ is to simply look at the periodogram and estimate the frequency of maximum amplitude of the Gaussian envelope of the peaks (Balona 2014).

High-speed photometry is widely used in asteroseismology, particularly
for the high-frequency pulsations present in white dwarfs, sdBV 
and roAp stars.  High-speed photometry does not refer to high-time
resolution, but rather that the target star varies more rapidly than the 
time for an observer to move the telescope between the  target and comparison 
stars. In such  a situation one should continuously observe the target star 
instead of observing the comparison stars. This technique is also known as 
non-differential photometry. To study stellar pulsation it is necessary to 
use integration time shorter than the pulsation period so that one can collect 
as many as data points over one pulsation cycle. Photometric integration times for asteroseismic targets are often selected to be 10\,-sec, and may be averaged to longer 
integrations. Differential photometry is the standard method for all 
photometric asteroseismic observations. In differential photometry it is always desirable to observe one or two non-variable comparison stars  of 
similar magnitude and color close to the target star to  correct
for the sky transparency variations.

Ground-based photometry normally consists of many CCD exposures of the target 
field.   The raw CCD data needs to be corrected for bias and flat fielding.
Synthetic aperture photometry of the target star and several comparison
stars in the same CCD frame are used to calculate the relative brightness
of the target with respect to comparison stars.  Sometimes
it is possible to increase the photometric accuracy by using profile fitting
to the stellar images on the CCD instead of aperture photometry (Stetson 1987). Profile fitting gives superior results in crowded fields or for very faint objects.
 
Spectroscopic observations of pulsating stars can provide additional
information particularly for the determination of the projected
rotational velocity, $v \sin i$. Surface rotation can also be used as an observable
to determine the age of the star. Spectroscopic studies of hundreds of stars in each spectral type from O to M has shown over the years that the early-type stars of classes O, B and A have rotational velocities between 200 and 350 km/sec, and that at spectral type F, there is a rapid decline from about 100 km/sec at F0 to 10 km/sec at G0. The Sun, a G2 star rotates at about 2 km/sec, and red giants rotate at 1 km/sec and slower.  Since stellar rotation slows with age, hence the rotation period of any star can be used to derive its age -- a technique referred to as ``gyrochronology'' which is an empirical relationship between rotation period, color, and age (Chanam´e \& Ram´ırez 2012). It provides means
by which surface rotation can be used to infer ages of cool stars
(e.g. Barnes 2009). The rotation-age relationship can be established but also for stars like our Sun. 

Stars are rotating and their rotation affects their oscillation frequencies. The information is relatively easy to conclude that the star rotates slowly provided observations are of superior quality that allows us to achieve the necessary high-accuracy of frequency measurements.
On the other hand, when the star is rotating fast, one should properly take into
account the consequences of the spheroidal shape of the star on the oscillation
frequencies before extracting seismological information on the internal structure
or the rotation profile of the star. The asymptotic relations works well when the 
rotation is small and the measurements of the rotational splittings is used to determine the internal rotation velocity.

Solar-type stars are generally slow rotators (in most cases
$v sin i < 20 $ km/sec) and the influence of rotation on the oscillation
frequencies is well known. However, distortion due to the
centrifugal force can have a large impact on the oscillation frequencies
even for slow rotators (e.g. Goupil 2009; Reese 2010,
and references therein). Such an effect is stronger for acoustic
($p$) modes with small inertia, which have a higher sensitivity
to the outer layers of the star. Therefore, their frequencies are
more sensitive to the physical properties of the surface, where
the centrifugal force becomes more effcient (Su{\'a}rez et al.
2010; Ouazzani \& Goupil 2012).  Much more evolved star such as a Cepheid is a radially pulsating star and
a slow rotator. However period ratios of radial modes can be quite significantly
affected by rotation as mentioned by Pamyatnykh (2003, Fig.6) for a $\delta$  Scuti
star and quantified by Su{\'a}rez et al. (2007) for a Cepheid. For hotter stars, rotation is more rapid and it is no longer possible to use the asymptotic relation. In fact, rotation  is the most serious obstacle to the interpretation of pulsating MS stars such as $\delta$~Scuti, SPB and $\beta$~Cep variables. An extensive review on the effect of rotation on $p$-mode pulsation has been given by  Goupil (2011). 

\section{Asteroseismic Modeling}

Reliable calculation of stellar models is the main objective of the asteroseismic 
investigations.  The pulsation properties are
sensitive to small effects and it is necessary to use the best possible
input physics when calculating a stellar model. It is necessary to
include additional effects such as convective overshoot (Di Mauro et al. 2003a, 2003b), 
as well as diffusion and settling of helium and heavy elements (Vauclair et al. 1974). 
In most models the Coriolis and the centrifugal forces induced by rotation are 
generally neglected.  When rotational is taken into account, it is usually
done, only to first order where perturbation theory can be used.  In case of solar-type stars, detailed classical models have been produced on several well observed targets viz. $\alpha$
Cen (Guenther \& Demarque 2000; Morel et al. 2000, Th{\'e}evenin et al. 2002; Eggenberger et al. 2004); $\eta$ Boo (Christensen-Dalsgaard et al. 1995; Di Mauro et al. 2004; Guenther 2004). A recent review on the problem and prospects of the modeling of the solar-like oscillators is presented by Di Mauro (2013).

A typical procedure of seismic modeling for the classical pulsators are as follow :

\begin{enumerate}

\item

The pulsational modes are identified in terms of the quantum numbers ($n,l,m$) 
that describe geometry of non-radial pulsations. The identification of these 
modes are the first step of the seismic modeling. 

\item

Photometric and/or spectroscopic time-series observations are used to
extract the frequencies, amplitudes and phases of the pulsations and to
identify the modes whenever possible. 

\item 

Evolutionary models which match the estimated physical parameters of the
target stars are calculated.  A pulsation model is used to calculate the
expected frequencies which are then compared to the observed frequencies with
known mode identifications.  The physical parameters of the evolutionary 
model are adjusted until a best match with observed frequencies is obtained.

\item

Once the best model is selected, the remaining frequencies without mode
identification can be identified by the matching them with the predicted
pulsation frequencies.

\end{enumerate}

\section{Pulsation Across the HR diagram}

The location of the major classes of pulsating stars are illustrated in  Fig. \ref{HR} and their pulsational characteristics  are listed in Table \ref{class}. The shading in this figure 
describes the excitation and the type of modes; (a) horizontal shading indicates stochastically excited $p$-modes, (b) NW-SE shading (like Cepheids) indicates self-excited (predominately) $p$-modes and (c) SW-NE shading (like DAV) indicates self-excited $g$-modes.

One can broadly divide pulsating variables into three groups : The first 
group contains the large amplitude Cepheid, RR Lyrae and cool red variables which
pulsate only in one or two radial modes. This group of stars follow a period-luminosity (PL) relation or a period-luminosity-colour (PLC) relation, making them useful as distance
indicators (McNamara, 1995; Petersen \& Christensen-Dalsgaard 1999;). The second group has low-amplitude oscillations in the mmag range  exhibiting many radial and non-radial modes and includes the white dwarfs, $\delta$ Scuti stars, roAp stars, 
$\beta$ Cephei stars, SPB stars and the $\gamma$ Dor stars. The third group 
has oscillations amplitude with amplitudes measured in $\mu$-mag and known as the solar-like oscillators.

\begin{table}
\caption{The names and basic pulsational characteristics  of pulsating variables. In the table the term F is refers for fundamental and O is for overtones. } 
\vspace{0.10cm}
\begin{tabular}{|c|c|c|c|c|} 
\hline
$Class$ &$Other$ $names$&$Mode$&$Period$&$Amplitudes$ \\
&&$Type$&$Ranges$&$(Light~variation)$ \\
\hline
Solar-like&main-sequence&$p$&3 to 10 min& $<$8 ppm\\
pulsator&red giants, sub-giants &$p$&few hrs to few days &few 10 ppm\\
        
\hline
 $\gamma$ Dor&slowly pulsating F &$g$& 0.3 to 3 d& $<50$ mmag  \\
\hline
 $\delta$ Sct&SX Phe(Pop.II)&$p$&18 min to 8 h&$<0.3$ mag\\
\hline
 roAp&$-$&$p$&5.7 to 23.6 min& $<10$ mmag    \\
\hline
 SPB&5 Per&$g$&0.5 to 5 d& $<50$ mmag    \\
\hline
 $\beta$Cep&$\beta$CMa,$\zeta$Oph&$p\& g$&2 to 8 h($p$)& $<0.1$ mag       \\
 $ $&53 per& & few days($g$)&$<0.01$ mag       \\  
\hline
 pulsating~Be&$\lambda$ Eri,SPBe &$p\& g$& 0.1 to 5 d & $<20$ mmag       \\
\hline
 pre-MS~pulsator&pulsating T Tauri,& $p$&1 to 8 h& $<5$ mmag      \\
  ~ & Herbig Ae/Be,& $p$&1 to 8 h& $<5$ mmag      \\
  ~ & T Tauri&$g$& 8 h to 5 d& $<5$ mmag      \\   
\hline
 $p$-mode sdBV&EC14026, V361Hya&$p$& 90 to 600 sec& $<0.3$ mag     \\
\hline
 $g$-mode~sdBV&PG1716+426&$g$&0.5 to 3 h& $<0.01$ mag       \\
 \hline
 $p$-mode~sdOV&&$p$&60 to 120 sec& $<0.2$ mag       \\
  \hline
PNNV&ZZLep& $g$&5 h to 5 d& $< 0.3$ mag      \\
\hline
DOV &, GW Vir& $g$&5 to 80 min& $<0.2$ mag     \\
\hline
DBV&V777Her&$g$&2 to 16 min& $< 0.2$ mag           \\
\hline
DAV&ZZCeti&$g$& 1 to 30 min& $<0.3$ mag          \\
\hline
 RR~Lyr&RRab& F& $\sim$ 0.5 d& $<1.5$ mag           \\
   $ $ &RRc&FO& $\sim$ 0.3 d& $<0.5$ mag         \\
    $ $&RRd&F+FO& 0.3 to 0.5 d& $<0.2$ mag           \\
  \hline
Type II Cepheid & W Vir& F& 0.8 to 35 d &$<1$ mag           \\
 & BL Her& F& 1 to 8d& $<1$~mag           \\
\hline
RV Tauri & RVa,RVb& F?& 30 to 150 d& $<3$~mag           \\
\hline
Type I Cepheid&Classical Cepheids &F&   $1$ to 135 d   &   $<2$ mag          \\
~ & s-Cepheid& FO& $<20$d& $<0.1$ mag     \\
\hline
 Mira&SRa,~SRb & $l=0$ & $>80$ d & $<8$ mag  \\
   & SRc & $l=0$& $>80$d&$<1$ mag \\
     & SRd& $l=0$& $<80$d&$<1$ mag  \\
\hline
\end{tabular}
\label{class}
\end{table}

Solar-like oscillations in main-sequence, sub-giants and giants are generally not visible from the ground except in a few cases. In the following subsections we give a brief introduction to the major lasses of pulsating variables.

\subsection{Rapidly Oscillating Ap Stars}

The rapidly oscillating Ap (roAp) variables  are sub-group of CP A-type cool ($T_{eff} \sim  6400 - 8500 K$) stars. These are  hydrogen core burning 
MS stars of mass around 2 M$_\odot$ and exhibit strong dipole magnetic 
field of the order of a few kG. The peculiarity in these stars results from 
atomic diffusion, a physical process common in all the CP stars having extremely 
stable atmospheres. The roAp stars pulsate in period ranging from 5.7--23.6 min 
characterized as high-order ($n>$20), non-radial and low-degree $p$-modes (almost pure 
dipole $l$=1). Typical amplitude of light variations lie in the range of 0.5-15 
mmag and radial velocity variation amplitude are of about 0.05-5 km$s^{-1}$. Most 
of roAp stars are multi-periodic, non-radial pulsators which makes them key objects for 
asteroseismology. 

Currently about 60 roAp stars are known and majority of them belong to the Southern 
hemisphere. The observed pulsation properties of roAp stars can be explained in 
terms of the oblique pulsator model (Kurtz 1982) in which the pulsation and 
magnetic axes are mutually aligned but tilted with respect to the rotation axis.  This 
model was modified by Dziembowski \& Goode (1996) and Bigot \& Dziembowski (2002) who found 
that the pulsation axis is not in exact alignment with the magnetic 
axes. Fig. \ref{hd12098lc} and \ref{hr1217-lc} show the sample light curves of two 
roAp stars and their corresponding amplitude spectrum are shown in Fig. 
\ref{hd12098ft} and \ref{hr1217-ft}. The excitation mechanism for the roAp stars is 
still an unresolved problem, although extensively debated over the years. In 
the H-R diagram they overlaps to the $\delta$ Scuti star instability strip (see 
Fig. \ref{HR}).  It is thought that the strong magnetic field suppresses
convection at the magnetic poles which drives the pulsation with high-radial overtones by the $\kappa$ mechanism in the H-ionization zone.
  
Space missions such as $\it Kepler$ has significantly improved our understanding of roAp stars  and at the same time additional
problems such as the unexpected appearance of low frequency modes have added (Balona et
al 2011) and the possible detection of modes with different axes of
pulsation (Kurtz et al. 2011).

Spectroscopy has now become an important tool for the detection of new roAp 
stars (Hatzes \& Mkrtichian 2004; Elkin et al. 2005; Kochukhov et al.
2009).  The advantages of spectroscopic observations over photometry is that
the roAp stars have particularly high-radial velocity amplitudes in the
lines of certain elements which are formed high in the atmosphere of the
star.  Several high-dispersion spectra taken over a few nights is sufficient 
to detect roAp oscillations which otherwise would have taken several weeks of
photometry to detect.  Such spectroscopic observations offer an unique 
opportunity to map the vertical structure of the pulsation modes 
(Ryabchikova et al. 2007) and for investigating the physics of propagating 
magneto-acoustic waves (Khomenko \& Kochukhov 2009). The peculiar atmospheres 
of magnetic roAp stars offer the building of a 3-D 
model of a pulsating stellar atmosphere (Kochukhov 2004). The roAp stars have also been observed interferometrically.  The first 
detailed interferometric study of the roAp star was applied on  $\alpha$ Cir for which allowed
a radius R = 1.967 $\pm$ 0.066 R$_\odot$ was derived (Brunt et al. 2010).

\subsection{$\delta$ Scuti Stars}

The $\delta$ Scuti stars are intermediate mass stars ($\sim$ 1.4 to 
3$M_{\odot}$) of spectral types A2--F5.  Their luminosity classes varies from 
III to V.  Most of $\delta$ Scuti stars belong to Population~I but some of  them show 
metallicities and space velocities typical to Population~II.  To date,
several thousands of $\delta$ Scutis stars have been found in our galaxy and these are among
the most common type of pulsating star (Breger 1979). 

Most of the $\delta$-Scuti stars are moderate or rapid rotators with surface 
velocities up to 100--200 km$s^{-1}$. The $\delta$ Sct stars pulsate in radial 
and/or non-radial low-order $p$ modes with periods in the range 18 min to 8 hr 
and amplitudes from mmag up to tenths of a magnitude. The non-radial pulsations 
found photometrically are low-overtones ($n$ = 0 to 7) and low-degree ($l\leq$ 3) 
$p$-modes. The radial pulsators mainly pulsate in the 
fundamental mode and its first few overtones.  For a typical $\delta$ Scuti
star,  $T_{\rm eff}$ = 7800 K, $M$ = 1.7 M${_\odot}$, $L$ = 15 $ L{_\odot}$, 
Y =0.28 and Z = 0.02 the pulsation constant, Q, and the ratio between 
periods is summarized in Table \ref{Q} (Breger 1979, Hareter et al. 2008). 

\begin{table}[!]
\caption{Pulsation constant, Period and Period ratios for a typical $\delta$ Scuti star.}
\centering
\begin{tabular}{c c c c c } 
\hline

Pulsation mode & Period & $P_i$/$P_{i-1}$& $P_i/P_F$ & Q (days) \\

\hline
Fundamental, F &  0.07861 & - &  1.000 &  0.0329 \\
1st Overtone, 1H &  0.05950 & 0.761 & 0.757&  0.0251 \\
2nd Overtone, 2H&  0.04846 &  0.810&  0.617&  0.0203 \\
3rd Overtone, 3H & 0.04095 & 0.845 & 0.521&  0.0172 \\
4th Overtone, 4H & 0.03533 & 0.862&  0.449 & - \\
5th Overtone, 5H & 0.03109 & 0.879&  0.396 & - \\
6th Overtone, 6H & 0.02774 & 0.882&  0.353 & - \\
\hline
\end{tabular}
\label{Q}
\end{table}

\begin{figure}
\center{
\includegraphics[width=1.1\textwidth,height=0.4\textheight]{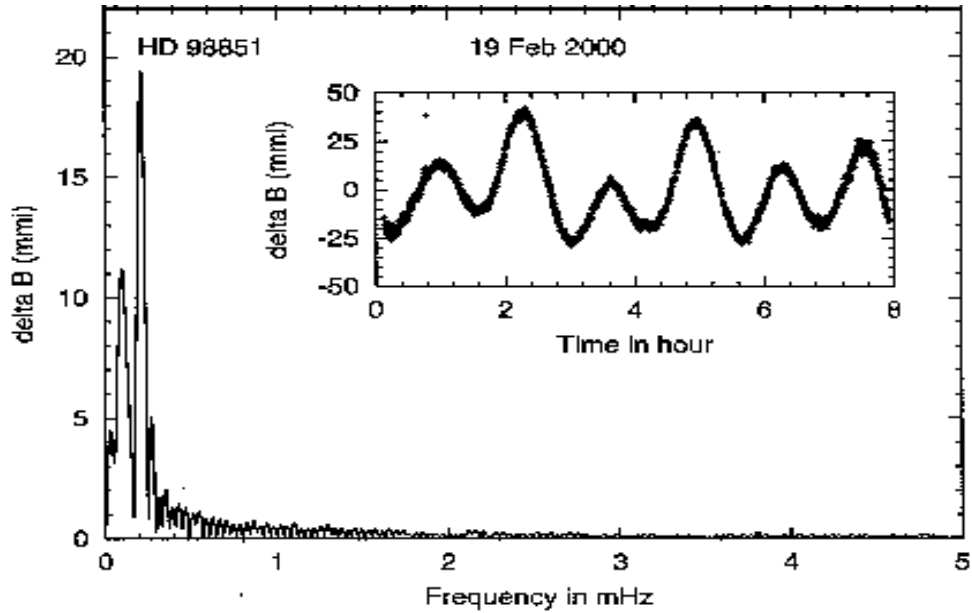}
}
\caption{: Light curve and amplitude spectrum of HD\,98851 observed from ARIES Nainital in 2000. 
Adopted from Joshi et al. (2003).}
\label{hd98851}
\end{figure}

Many $\delta$ Scuti stars are multi-periodic variables therefore they are good candidates 
for asteroseismology. For example 79 and 29 frequencies are detected in FG Vir 
and 44 Tau, respectively (Breger et al. 2005).  
Fig. \ref{hd98851} shows the light curve and amplitude spectrum of a typical 
$\delta$ Scuti star HD\,98851 discovered from ARIES Nainital under the ``Nainital-Cape 
Survey''. The $\delta$ Scuti stars are situated where the classical 
instability strip crosses the main sequence (see Fig. \ref{HR}) hence the 
excitation mechanism of pulsation in $\delta$ Scuti stars is the 
$\kappa$-mechanism, same as for other stars in the classical instability strip. 
The driving zone are {\it H\,I and He\,II} ionization zones which provides  enough 
counterbalance to the damping in the underlying layers (Breger 2000).

The identification of the oscillation modes is a very complex task for 
$\delta$ Scuti stars, since the asymptotic theory does not apply to the excited 
modes (low-order $p$-modes). Thus, at present, asteroseismology 
is able to put additional constraints on the internal structure of $\delta$ Scuti 
stars. The P-L relations for $\delta$ Scuti stars have been established in the V band 
(Petersen \& Christensen-Dalsgaard, 1999; Pych et al. 2001; Templeton et al. 2002), V and R band (Garg et al. 2010) and V and I-Wesenheit-index (Majaess et al. 2011). Using these
relations, one can determine the distance of these stars independently.

Thanks to the different space missions which provided asteroseismic data of 
$\delta$ Scuti stars with very high precision. Matthews (2007) found 88 frequencies 
in the $\delta$ Scuti star HD\,209775 observed by MOST.
The regular patterns were also found in the oscillation spectra of 
$\delta$ Scuti stars observed by $\it CoRoT$ and $\it Kepler$ (Garc{\'i}a  et al. 2013 ;
Mantegazza et al. 2012).

There are two well-defined sub-groups of $\delta$ Scuti stars : (1) High-amplitude $\delta$ 
Scuti stars (HADS) which are first classified as AI Velorum stars. These stars pulsate 
in the fundamental or first overtone modes with V-amplitude $\geq$ 0.3-mag and 
follow the P-L relation. Hence they have been used to estimate the distance of  
the LMC and to star clusters; (2) SX Phoenicies (SX Phe) are $\delta$ Scuti stars 
of Population II, with shorter periods and lower amplitudes. They have been found 
in globular clusters and are known as Blue Stragglers. From an evolutionary point of view they 
are unusual and understood resulted from merged binary stars (Mateo et al. 1990).

\begin{figure}
\center{
\includegraphics[width=0.9\textwidth,height=0.6\textheight]{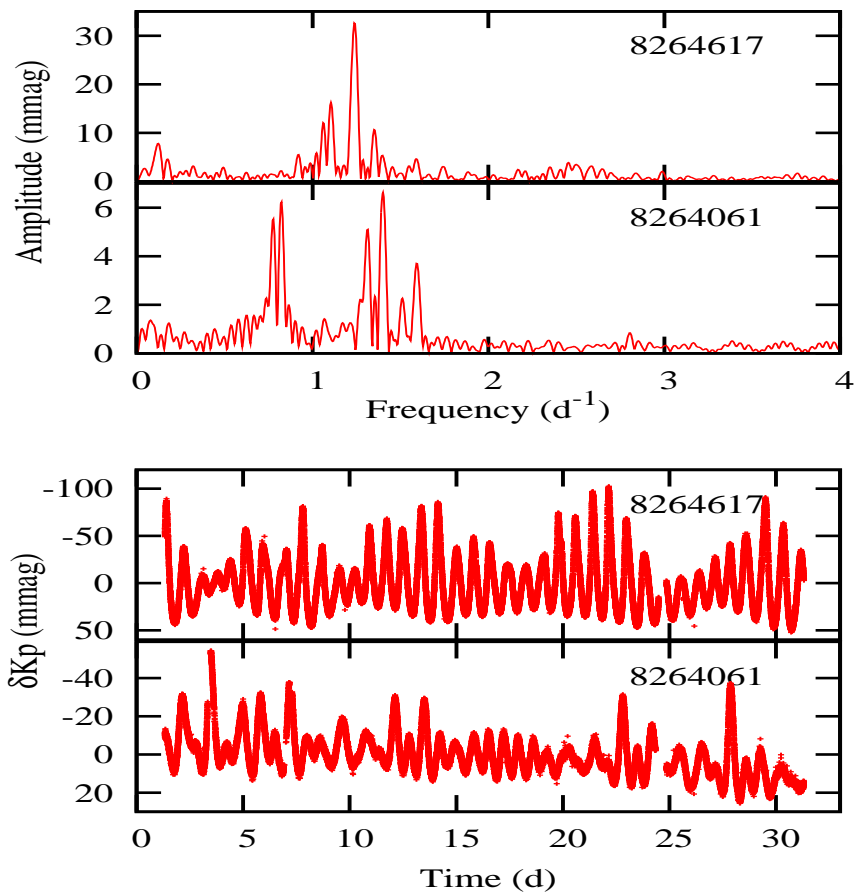}
}
\caption{: Light curve (bottom) and periodograms (top) of two $\gamma$ Dor stars KIC8264617 and KIC8264061 in the field of open star cluster NGC6866. Adopted from Balona et al. (2013).}
\label{hl}
\end{figure}

\subsection{$\gamma$ Dor Stars}

Balona et al. (1994) discovered a new group of Pop I stars known as $\gamma$ Dor 
which are the main-sequence stars that partly overlap the cool edge of the 
$\delta$ Scuti instability strip.  About 50 confirmed members of 
this class are known to the date and more than 100 additional candidates are being 
studied observationally (Henry et al. 2005; De Cat et al. 2006). The $\gamma$ Dor stars are 
early-type-F stars that have masses between 1.5 and 1.8$M_\odot$ and pulsate 
in multiple non-radial, $g$-modes with periods ranging from 0.3 to 3.0 d. The 
high-order $g$-modes are driven by a modulation of radiative flux from the 
interior of the star due to the convection zone, a mechanism known as convective 
blocking (Guzik et al. 2000)

Stars that exhibit both $p$-modes  and $g$-modes are very  important for the 
asteroseismic study because they pulsate with many simultaneous frequencies and 
can probe both the surface and core region of the stars. There is considerable 
overlap between the $\delta$ Sct and  $\gamma$ Dor instability strips where 
both the high-frequencies $\delta$ Sct and the low-frequencies $\gamma$ Dor 
stars are found. These stars are known as hybrid stars (Grigahc{\'e}ne et al. 2010; Antoci et al. 2011) and first discovered in the A9/F0V star HD\,209295 (Handler et al. 2002).

\subsection{RR Lyrae Stars}

The RR Lyra are evolved low-mass (M $\leq  7 M _{\odot}$) stars of spectral 
type A2-F6 on the horizontal branch (core He burning) with a low content of 
heavy elements and are commonly found in globular clusters and in 
the Magellanic Clouds. About 90 percent RR Lyrae  pulsates radially either 
in fundamental or first overtone modes with period 0.2 to 0.5 d having amplitude 
between 0.2 to 1.5 mag in V band. In the H-R diagram, they are located just below 
the Cepheids in the instability strip. Due to their  mono-periodicity, 
RR Lyrae are not suitable for the seismic study but they follow a period-luminosity-color relation, 
hence can be considered standard candles for distance measurements and galactic 
evolution (Kolenberg et al. 2010).

\begin{figure}
\center{
\includegraphics[width=1.0\textwidth,height=0.2\textheight]{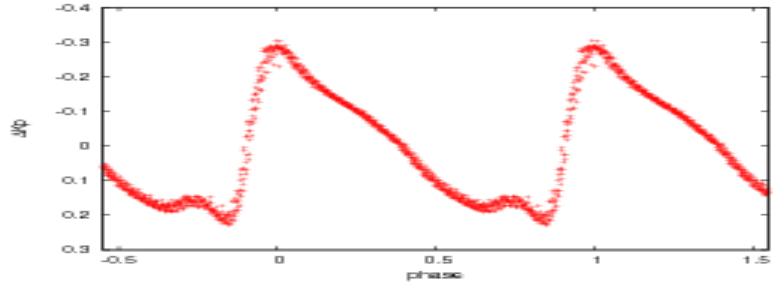}
}
\caption{: Phase light curve plot of RR Lyr at maximum and minimum amplitude. Adopted from Szab{\'o} (2010).}
\label{rrlyra}
\end{figure}

In one member of the class, Blazhko (1907) discovered that the maximum amplitude 
varied cyclically with a period of 40.8 d i.e. amplitude and phase modulation (Kolenberg 2008). 
Using $\it Kepler $ data, Szab{\'o} et al. 
(2010) found Blazhko effect in three  RR Lyrae stars which showed period doubling 
in certain phases of the Blazko cycle, with slight variations in
the maximum amplitude between alternating pulsation cycles. The two competing 
theoretical explanations for the Blazhko effect are :(1)  caused by the excitation 
of a non-radial oscillation mode of low-degree apart from  the main radial mode, 
through non-linear resonant mode coupling. In this model the Blazhko period is 
interpreted as the beat period between the radial fundamental and a non-radial 
mode (Dziembowski \& Cassisi 1999); (2) caused by a magnetic field that influences the oscillations similar 
to the oblique pulsator model for the roAp stars. In this case the Blazhko period 
is interpreted as the rotation period of the star (Takata \& Shibahashi 1995).

\subsection{Cepheids}
 
Cepheid are are population I yellow super-giants of spectral class F6 - K2. 
These are very luminous stars with luminosity 500 to 300,000 times more than the 
Sun, pulsate with periods from 1 to 135 days and light variations from 0.1 to 
2 mag. The light curve of one of typical $\delta$ Cephei is shown in Fig. \ref{cephi}. By measuring the oscillation period of a Cepheid 
and using the period-luminosity relation, one can derive the absolute magnitude, 
hence distance of the star. For this reason, Cepheids are also called distance 
indicators. While stellar parallax can only be used to measure distances to stars 
within hundreds of parsecs but Cepheids can be used to measure the larger distances 
of the galaxies where they belong. Today we recognize the following stars as 
distinct classes of Cepheids :

\begin{figure}
\center{
\includegraphics[width=1.0\textwidth,height=0.3\textheight]{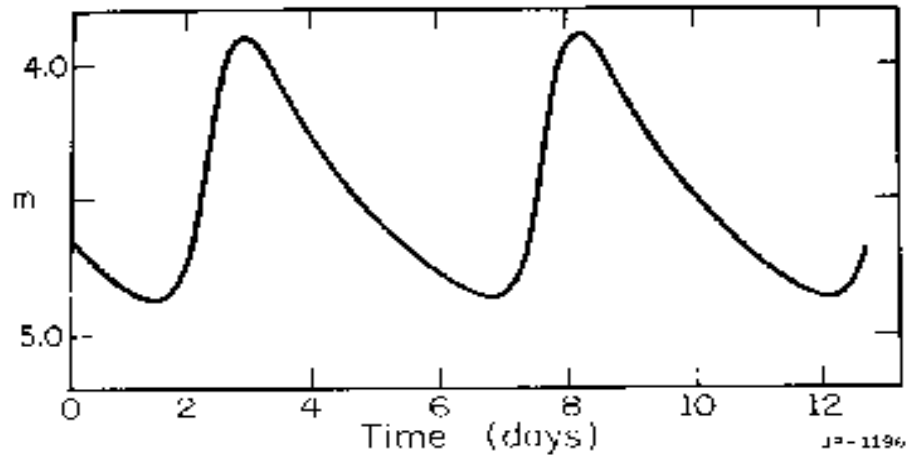}
}
\caption{: Typical light curve of a Cepheid.}
\label{cephi}
\end{figure}

\subsubsection{Classical Cepheids or Type I Cepheids}

These are massive (M $\sim$5--15$M_{\odot}$) young bright giants or super-giants of 
spectral types F or G and are found exclusively in the disk population of 
galaxies. The periods of this class lie in the range 1 $<$P$<$ 135 d, with 
amplitudes of 0.5-2 mag. Cepheids are core He burning stars and cross the 
instability strip up to three times.

\subsubsection{W Vir stars}

These are evolved F6--K2 Population II giants with periods in the range 
0.8 $<$ P $<$ 35 d and amplitudes of 0.3--1.2 mag. Their light curves generally 
resemble those of the classical Cepheids, particularly if they have a period 
of 3--10 d. Typically, they have lower masses than classical 
Cepheids are about 1.5 mag fainter than classical 
Cepheids of the same period and obey a different P-L relation. 
 
\subsubsection{BL Her stars}

These Cepheids have periods less than about 8 days and mostly are metal deficient 
and thought to be post-horizontal branch stars. They have a characteristic bump 
on the declining branch of the light curve.

\subsubsection{RV Tauri stars}

The RV Tauri are characterized by alternating deep and shallow minima in their 
light curves with periods 30--150 d measured between successive deep minima. 
These stars  have spectral types F--G at minimum and G--K at maximum. RV Tauris are 
probably low-mass stars in transition from the asymptotic giant branch to white dwarfs.

\subsection{Slowly Pulsating B Stars}

The Slowly Pulsating B Stars (SPBs) are situated along the MS indicating that nuclear core hydrogen burning is still ongoing. SPBs are massive hot stars with
spectral types ranging between B2 and B9, corresponding to effective temperatures
of 12000K up to 18000 K. Their masses lie in the range between 3$M_{\odot}$ to
7$M_{\odot}$. All the known members of this class are slow rotators in the sense that the observed projected rotational velocities are much smaller than the break-up
velocities of such stars. The SPB stars are mid-to-late B-type stars oscillating in high-order $g$-modes with periods from 0.3 to 3 days similar to the 53 Per stars. These oscillations are driven by the $\kappa$-mechanism operating  in the iron ionization zone located at $T_{eff} \sim $ 200000 K (Gautschy \& Saio 1993). Since most SPB stars are multi-periodic, hence the observed variations have long beat periods and are 
generally complex. 

\subsection{$\beta$ Cephei Stars}

The  $\beta$ Cep stars have been known as a group of young Population I near 
MS pulsating stars for more than a century. They have masses between 
8 and 18 $M_{\odot}$ and oscillate in low-order non-radial $p$- and $g$-modes 
with periods between 2 and 8 h with amplitudes less than 0.3-mag in V band 
(Moskalik 1995). More than 100 members of this group are known and the class 
contains dwarfs up to giants (Stankov \& Handler 2005). Most of the $\beta$ Cep 
stars show multi-periodic light and line profile variations (Pamyatnykh 1999), 
hence considered as excellent objects for the asteroseismolgy (Balona et al. 2011).

\subsection{Mira Variables}

Miras represent the advanced evolutionary stages of low and intermediate mass 
stars such as Sun. These are class of pulsating variables characterized by 
pulsation periods between 10 and 100 days with amplitudes greater than
2.5-mag in V-band. These are cool 
red giant and are found in the high-luminosity portion of the asymptotic giant 
branch (AGB) in the H-R diagram (Tabur et al. 2009). 
There is some dispute whether they pulsate primarily in their fundamental
or first overtone (Wood 1995), and there is evidence that some Miras switch 
between different modes on time scales of decades (Bedding et al. 1998).
 
\section{Sub-dwarf B Variable Stars}

\begin{figure}
\center{
\includegraphics[width=1.0\textwidth,height=0.2\textheight]{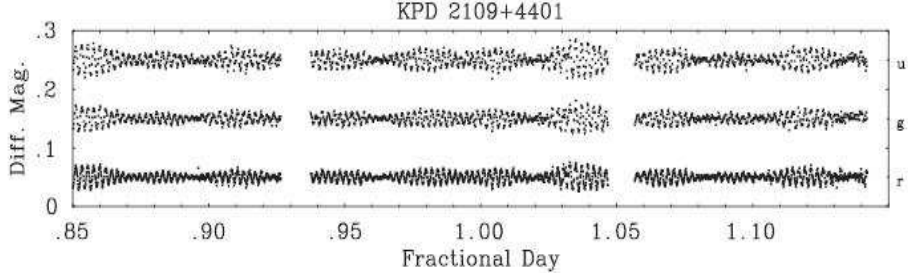}
}
\caption{: Partial ultracam light curves for sdBVs KPD 2109+4401. Adopted from Jeffery et al. (2004).}
\label{sdb}
\end{figure}

The  Sub-dwarf B (sdB) stars are core-helium burning stars of $\sim 0.5 M_\odot$ with a very 
thin surface layer of hydrogen. With effective temperatures 25000 $\leq
T_{\rm eff}/K \leq$ 35000, the atmospheres of these stars are entirely radiative. On the HR diagram, the sdB stars are found between the upper main 
sequence and the white dwarf sequence. Pulsations in 
sdB were first observed in such stars by Kilkenny et al. (1997) characterized 
as low-degree ($l$), non-radial multi-periodic $p$-mode pulsators with period 65-sec$\leq$P$\leq$500-sec, and amplitudes ranging from 1 mmag up to 0.3 mag. The driving 
arises from the $\kappa$-mechanism operated by  opacity from the iron-group 
elements. Three types of pulsating sub-dwarf star are known: long-period 
sub-dwarf B (sdB) stars (PG1716 or V1093 Herculis stars) that oscillate in 
high-overtone, $g$-modes with periods from 30 to 180 minutes, short-period sdB 
stars pulsating in low-overtone, $p$- and $g$-modes with periods from 90 to 600 
sec (EC14026 or V361 Hydrae stars), and the oscillating sub-dwarf O (sdO) 
stars are low-overtone $p$-mode pulsator with periods from 60 to 120 sec.
Fig. \ref{sdb} shows a portion of the light curve obtained with ULTRACAM a 
high-speed 3-channel CCD camera mounted on the 4.2 m William Herschel telescope. 
At ARIES we are developing a similar three channel fast CCD photometer for the 
3.6-m telescope (see Sec. 15). 

\subsection{Solar-like Oscillations}

The well known solar-like oscillator is of course the Sun and it's frequency spectrum 
is shown in Fig. \ref{solarspec} where one can see hundreds of peaks centered at 
3 mHz (P=5 min). The solar-like oscillation can be expected in low-mass main-sequence stars, sub-giants, stars on the red-giant branch (RGB) horizontal branch and asymptotic-giant branch (Christensen-Dalsgaard \& Frandsen 1983; Houdek et al. 1999, Dziembowski et al. 2001). The oscillation periods in these stars are expected to be in the range from the typical 5-min (MS stars), as in the Sun, up to about a few days in sub-giant and giant stars. According to theoretical estimations (Kjeldsen \& Bedding 1995) the expected luminosity variations are of the order of a few mmag in giant stars while in MS
and sub-giant stars are of the order of ppm which is below the detection limit for
ground-based photometric observations. The radial velocity
amplitudes are expected to be atmost 50-60 m/sec
in K giants, 1-2 m/sec in F and G sub-giants and even
smaller in main-sequence solar-type stars. 

\begin{figure}
\center{
\includegraphics[width=0.9\textwidth,height=0.6\textheight]{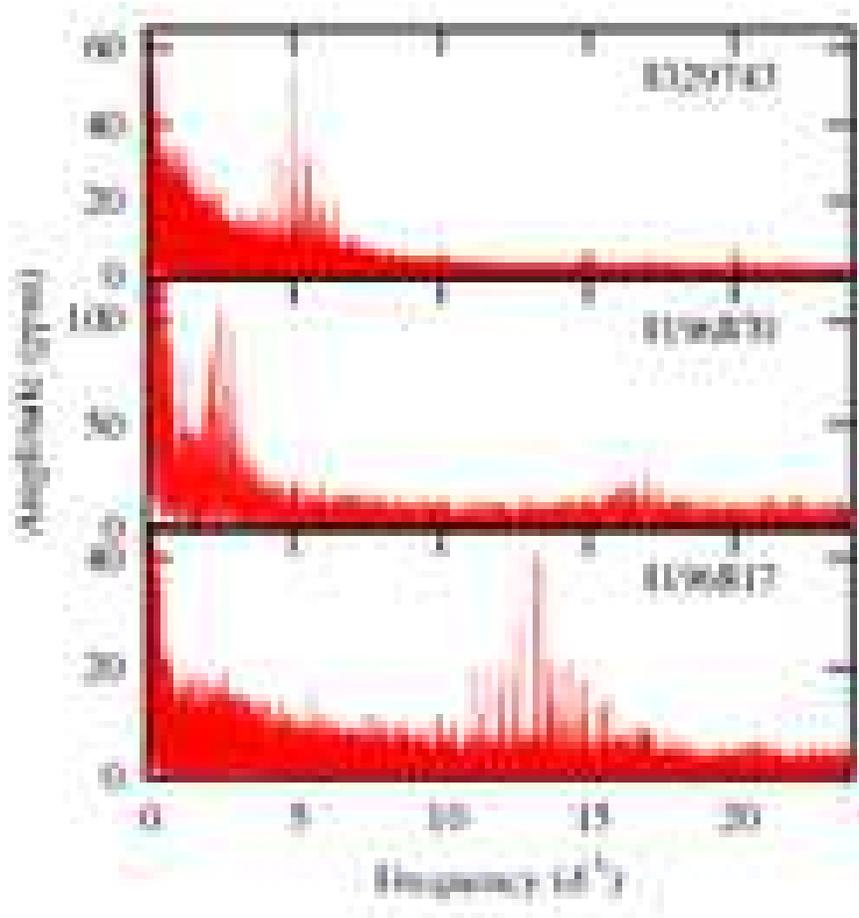}
}
\caption{: Periodograms of three solar-like oscillators the field of open star cluster 
NGC\,6866 observe by $\it Kepler$. Adopted from Balona et al. (2013).}
\label{sol}
\end{figure}

The observed modes of solar-like oscillators are typically high-order non-radial, 
acoustic $p$ -modes.  However a distant observer can usually not resolve the stellar
surface and can only measure the joint effect of the pulsations in light and radial
velocity. As a consequence, the effects of high-spherical degree oscillations average
out in disk-integrated measurements (Dziembowski
1977) and due to this only modes of low-angular degree ($l$) can be observed. 
A revolution in the detection of solar-like oscillation took place  photometric 
observations after the launch of $\it MOST, CoRoT$ and $\it Kepler $ space missions. The $\it Kepler $ has yielded clear evidence of solar-like oscillations in more than 500 solar-type stars (Chaplin et al. 2011). 

 The simplest analysis of asteroseismic data  of the solar-like oscillators is  based on the overall properties of the oscillations.  In the periodogram, solar-like oscillations are easily identified because of the localized comb-like structure where amplitudes decrease sharply from a central maximum. Fig. 15 shows an example of comb-like structure generally found in solar-like oscillators.  
structure. The frequency of maximum amplitude ($\nu_{\rm max}$) and large separation (${\Delta \nu} $) between successive overtones of the same degree depends on the mean density of the star (Ulrich 1986; Kjeldsen \& Bedding 1995) and expressed as :

\begin{equation}
\frac{\nu_{\rm max}}{\nu_{\rm max}\odot} \approx \frac{M/M_\odot}{(R/R_\odot)^2
\sqrt{T_{\rm eff}/T_{{\rm eff}\odot}}},
\end{equation}

where $T_{\rm{eff},\odot}$ = 5777 K, $\nu_{{\rm max}\odot} = 3120$\,$\mu$Hz (Kallinger et al. 2010), $M/M_\odot$, $R/R_\odot$ and $T_{\rm eff}/T_{{\rm eff}\odot}$ is the stellar mass,
radius and effective temperature relative to the Sun.

\begin{equation}
\frac{\Delta \nu}{\Delta \nu_{\rm \odot}} \approx \frac{(M/M_{\rm \odot})^{1/2}}{(R/R_{\rm  
\odot)})^{-3/2}}
\end{equation}

The above scaling relations can be easily expressed in the terms of  stellar  radius $R$ and 
mass $M$ as function of $\delta\nu$, $\nu_{max}$ and effective temperature  $T_{\rm{eff}}$:

\begin{eqnarray}
\frac{R}{R_\odot} & = & \left(\frac{\nu_{\rm{max}}}{\nu_{\rm{max},\odot}}\right)
 \left(\frac{\delta\nu}{\delta\nu_\odot}\right)^{-2} \left(\frac{T_{\rm{eff}}}{T
_{\rm{eff},\odot}}\right)^{1/2}\\
\frac{M}{M_\odot} & = & \left(\frac{\nu_{\rm{max}}}{\nu_{\rm{max},\odot}}\right)
^3 \left(\frac{\delta\nu}{\delta\nu_\odot}\right)^{-4} \left(\frac{T_{\rm{eff}}}
{T_{\rm{eff},\odot}}\right)^{3/2}
\end{eqnarray}

Thus, provided that $T_{\rm{eff}}$ is known, both radius and 
mass of solar like oscillations can be easily derive from the frequency  spectrum. This technique has been applied to huge samples of MS, sub-giant and red-giant stars (Bedding \& Kjeldsen 2003; Chaplin 2011a,b; Huber 2011; Stello et al. 2008; 
White et al. 2011; Miglio et al. 2012; Hekker et al. 2013). Efforts have recently been taken to test the accuracy of the scaling relations and asteroseismically inferred properties. The reliability of the mass and  radius estimates from these relations depends on the validity of the scaling laws themselves. For a comprehensive review on solar-like oscillations see Chaplin \& Miglio (2013) and references therein.

\subsection{Planetary Nebula Nuclei Variables}

The Planetary nebula nuclei variables(PNNV) are pre-white dwarf stars situated at the 
center of planetary nebula. These are multi-periodic variable pulsate with period 
10--35 min with amplitudes of around 0.1 mag.

\subsection{White Dwarfs}

White dwarfs are the final evolutionary stage of about 98\% of stars. A fraction of them are $g$-mode pulsators with hydrogen atmospheres and are located in a narrow 
range of effective temperature 10500 $< T_{eff}  < 12300 $K (Winget \& Kepler 2008; Althaus et al. 2010). In fact other than 
the Sun white dwarfs  represent the stars in which the largest number of oscillation 
frequencies have been detected. When the white dwarfs evolve 
and pass through the instability strip they become pulsators and known as the DAV (ZZ Ceti), 
DBV (V777 Her) and DOV (GWVir) (D = white dwarf; V = pulsating variable; and O(hot), B(warm)  and A(cool) refer for spectral type. The ZZ Ceti (or DAV) stars are the most numerous class of degenerate white dwarfs, with $\sim$ 160 members known to date (Castanheira et al. 2013). Their photometric variations are characterized as non-radial $g$-mode pulsations with low-harmonic 
degree ($l \leq $ 2) of periods range 70 to 2000 sec with amplitude variations 
up to 0.3 mag. 

The driving mechanism thought to excite the pulsation near the blue edge of 
the instability strip is the $\kappa-\gamma$ mechanism acting on the hydrogen 
partial ionization zone (Winget et al. 1982). 
Fig. \ref{hl} shows a typical light curve of a DAV white dwarf HL Tau 76 observed 
from ARIES Nainital under a multi-site campaign. 

\begin{figure}
\center{
\includegraphics[width=0.9\textwidth,height=0.5\textheight]{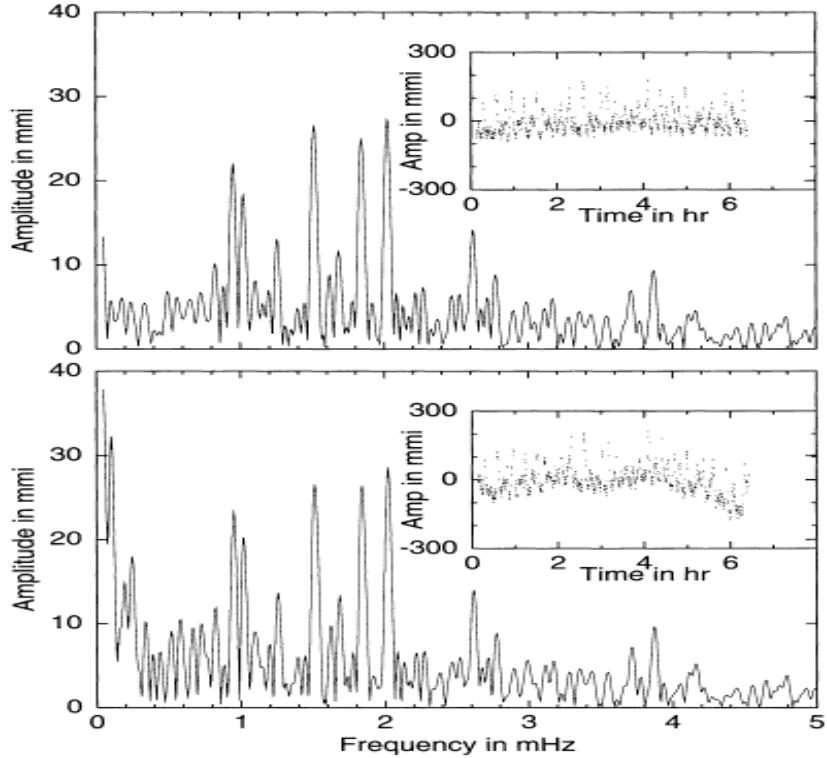}
}
\caption{Light curve of a DAV (ZZ Ceti) white dwarf observed from ARIES in 1999. Adopted from Ashoka et al. (2001).}
\label{hl}
\end{figure}

White dwarfs asteroseismology acts the comparison between the observed pulsation 
periods and the periods computed for appropriate theoretical models to place the 
observational constraints on their stellar mass, the thickness of the outer 
envelopes, the core chemical composition, magnetic fields, rotation rates and 
crystallization process (Montgomery \& Winget 1999; Metcalfe et al. 2004; Corsico et al. 2005; 
Kanaan et al. 2005).  The rate of period change can be 
employed to measure their cooling rate (Kepler et al. 2005). Many attempts have 
been undertaken to derive the internal rotational profiles by using the observed 
splitting and adopting forward calculations. The inversion techniques for white dwarfs have been presented in several works by Kawaler et al. (1999) and Vauclair et al. (2002).

\section{Asteroseismology of Clusters}

Stars in clusters are believed to formed from the
same cloud of gas roughly at the same time, hence the cluster members
are therefore expected to have common properties.
Asteroseismology of clusters is a potentially powerful tool to test
aspects of stellar evolution that cannot be addressed otherwise. The advantage of cluster asteroseismology is that the seismic data do not suffer from uncertainties in distance or extinction and reddening and for this reason, ensemble asteroseismology
is suitable for cluster stars (Stello et al. 2010, 2011a,b;  Miglio et al. 2012; Chaplin et al. 2014; Wu et al. 2014a).

To bring a new dimension to ensemble asteroseismology one need high duty observations cycle and with a long time base of at least half a year to detect rotational splitting of the pulsation modes of numerous cluster stars, including the slowest rotators. Prior to $\it Kepler $ survey many attempts were undertaken to detect solar-like oscillations in open and globular clusters (Gilliland et al. 1993; Stello et al. 2007a,b; Stello \& Gilliland 2009). Stello et al. (2010) obtained the clear evidence of the detections of solar-like oscillations in red giants of open cluster NGC6819 and were able to measured the large frequency separation ($\Delta \nu$), and the frequency of maximum oscillation power ($\nu_{max}$).

\begin{figure}
\center{
\includegraphics[width=0.8\textwidth,height=0.4\textheight]{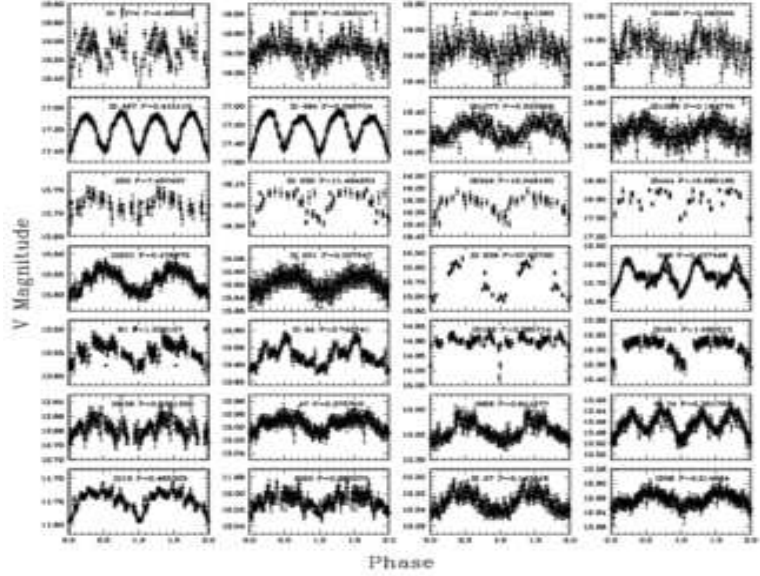}}
\caption{ : V-band phased light curve for the 28 variable stars identified in this study. Adopted from Joshi et al. (2012a)}
\label{light}
\end{figure}

Distance is a fundamental parameter in astrophysics which can be derived using the data from Hipparcos satellite (Perryman \& ESA 1997), binary systems (Brogaard et al. 2011), isochrone fitting (Bedin et al. 2008; Hole et al. 2009; Wu et al. 2014a), red-clump stars as standard candles (Garnavich et al. 1994; Gao \& Chen 2012), the P-L relation of pulsating stars (Soszynski et al. 2010). Wu et al. (2014b) derived a new relation for distance of two clusters NGC 6791 and NGC 6819 by measuring the global oscillation parameters $\Delta \nu$  and $\nu_{max}$. Stello et al. (2010) performed an asteroseismic analysis based on the first month of data from the $\it Kepler $ Mission to infer the cluster membership for a small sample of red giant stars in NGC 6819. The authors demonstrated that cluster membership determined  from seismology show more advantages over other methods.

\begin{figure}
\center{
\includegraphics[width=0.8\textwidth,height=0.4\textheight]{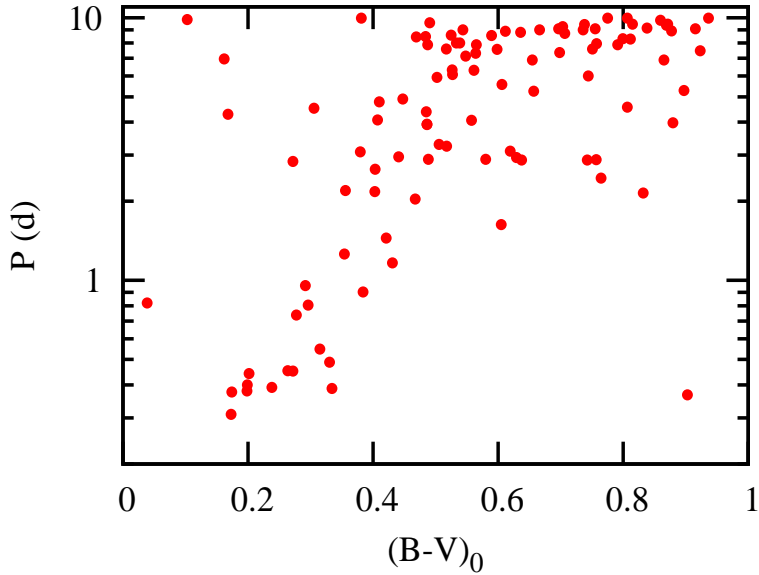}
}
\caption{: Rotation period P (in d) as a function of $(B-V)_0$ for main-sequence
stars in the field. Adopted from Balona et al. (2013)}
\label{rot}
\end{figure}

In order to investigate the occurrence of pulsation in A- and F-type stars in 
different galactic environments, at ARIES we initiated a survey to 
search for the photometric variability in the young and intermediate age open 
star clusters. We selected sample of clusters having  different age 
(young to intermediate of age range 10 to 100 Myrs). The preliminary 
results of the survey resulted 18 new variables in an open star cluster NGC6866 
(Joshi et al. 2012a). The V-band phased light curves of 28 variables are
shown in Fig. \ref{light}.

This cluster belongs to the $\it Kepler $ field, therefore an analysis of stars in the field of the open cluster NGC 6866 was performed time-series photometry from the $\it Kepler $ data. We identified 31 $\delta$ Scutis,
8 $\gamma$ Dor pulsating variables and 23 red giants with solar-like oscillations. There are 4 eclipsing binaries and 106 stars showing rotational modulation with indication of star-spots. We attempted to identify cluster members using  their proper motions but found very poor discrimination between members and non-members (Balona et al. 2013). 
 
The rotation periods of the MS stars are correlated with colour, so that a period--age--mass relation can be derived from open clusters and applied to stars of unknown ages.  We noticed that the correlation applies not only to cool stars but extends up-to A-type stars in the cluster which is shown in Fig. \ref{rot}.  We conclude that either the role of convection in A-type stars is not fully appreciated or that something other than convection is at the root of the rotation period-mass-age relationship (Balona et al. 2013). 

\section{ARIES Contribution towards the Asteroseismology}

ARIES is contributing significantly in the area of asteroseismology since last two decades. In the following sub-sections we high-light the programmes where we participated for the asteroseismic study of pulsating variables. 

\subsection{The Nainital-Cape Survey}

Aiming to search for the new roAp stars in the northern hemisphere a survey ``The Nainital-Cape Survey'' is being carried out at ARIES since year 1999. The high-speed/fast photometric technique was adopted for the survey. Each star was observed through a Johnson B filter with continuous 10-sec integration. The observations were acquired in a  single-channel with occasional interruptions to measure the sky background, depending on the phase and position of the moon. To minimize the  effects  of seeing  fluctuations  and  tracking  errors an  aperture  of 30$^{\prime\prime}$ was selected. Each target was observed continuously for  1 to 3 hours a sufficient time to revel the roAp like oscillations.The results obtained from this survey are described by Martinez et al. (2001); Girish et al. (2001); Joshi et al. (2003; 2006; 2009; 2012b). 

\begin{figure}
\center{
\includegraphics[width=1.0\textwidth,height=0.2\textheight]{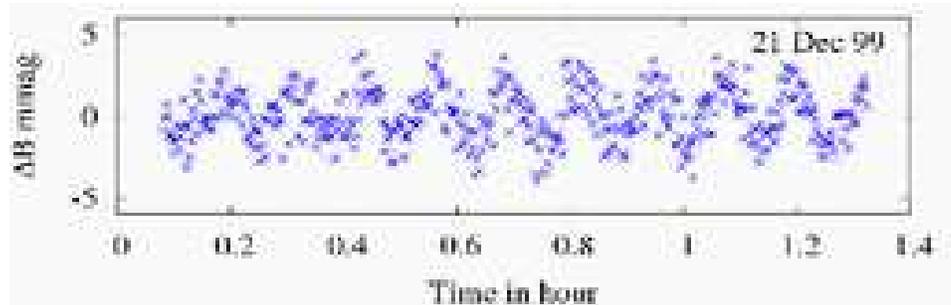}
}
\caption{: Light curve of HD12098 observed on the night of 21 December 1999.
Each dot represents a 10-s integration. Adopted  from Girish et al. (2001).}
\label{hd12098lc}
\end{figure}

\begin{figure}
\center{
\includegraphics[width=1.0\textwidth,height=0.4\textheight]{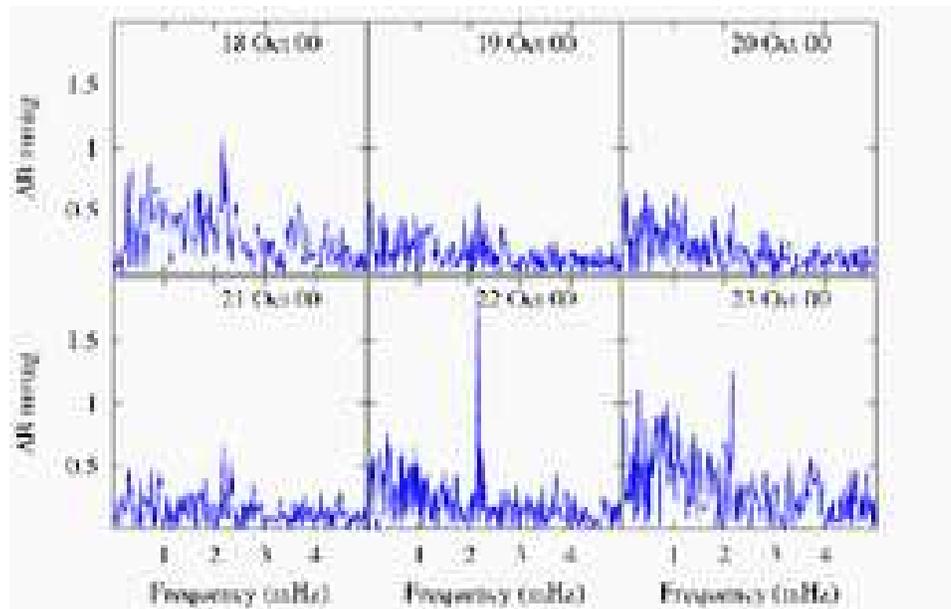}
}
\caption{: Amplitude spectrum of the time-series of HD\,12098  obtained on different nights. Adopted from Girish et al. (2001).}
\label{hd12098ft}
\end{figure}

Based on the Str\"{o}mgren colours we searched for the rapid oscillations in HD\,12098 on the night of 21 November 1999 (HJD 2451504). We were rewarded with the discovery of 7.6-min oscillation which was confirmed on the nights of HJD 2451505, 2451534
and 2451535. Fig. \ref{hd12098lc} shows the light curve obtained on
night of HJD\,2451535 and  Fig. \ref{hd12098ft} shows the amplitude spectra of the light
curves obtained during the follow-up observations. The analysis of data of individual night clearly shows the modulation in the amplitude of oscillation. 

A multi-site campaign was organized on HD\,12098 in Oct/Nov 2002 aiming to search for additional frequencies and derive the basic physical parameters. The campaign involved a total of  ten observatories and  of 394 hours of useful data extending over 28 nights with 45\% duty cycle were obtained. The equal separation of the frequencies seen in the multi-site data gives a rotation period of $\Omega$= 5.41$\pm$0.05 day for HD\,12098 which is very close to the 5.5 days rotation period predicted by Girish et al. (2001).

\subsection{WET Campaign on pulsating sdBV }

Astronomers at various observatories in India are involved in the observation of variable phenomena in astronomical objects where continuous observations are required.  The observational coverage from the Asian longitudes are very important  to fill critical gaps in the data sets.  In order to take the advantage of the geographical location of India, we first joined the WET organization in November
1988 to observe V471 Tau and since then we have participated in seven campaigns. ARIES participated in the multi-site campaign organized for PG 1336-018 (NY Vir) which is a close eclipsing binary with a binary period of 2.4 hr (0.101d) and one of the components of this binary system is a pulsating B sub-dwarf (sdBV).  

\begin{figure}
\center{
\includegraphics[width=0.9\textwidth]{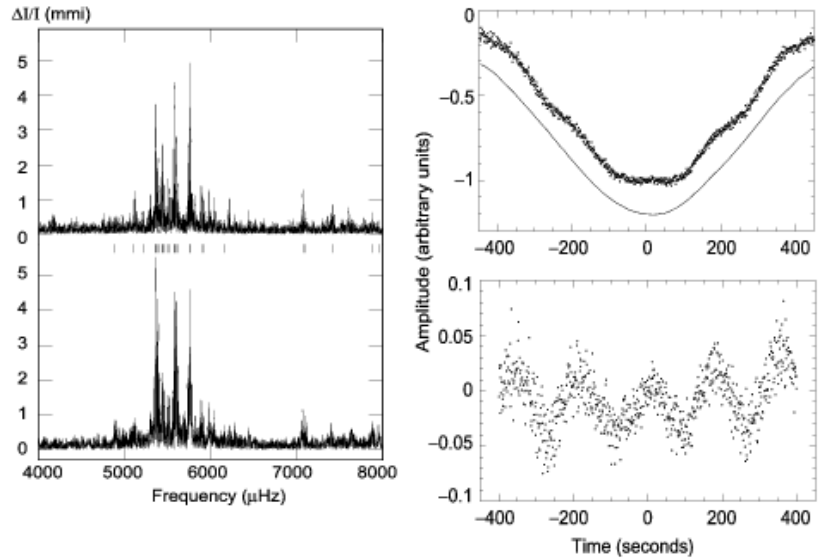}
}
\caption{: The figure on the left side shows the frequency spectrum of PG1336-018 obtained during two halves of the WET run in April 1999. The figure on the right side shows the presence of pulsations during the eclipse after removing the eclipse profile with best fit. Adopted from Kilkenny et al. (2003).}
\label{pg13}
\end{figure}

PG 1336-018 was observed in April 1999 under the WET campaign and 172 hours of observation with 47\% coverage was obtained.  The main aim of the campaign was to look for pulsations during eclipse and if well resolved, mode identification would be possible depending on the gradual decrease/increase in amplitude of the frequencies during ingress/egress. More
than 28 frequencies were detected with 20 frequencies were identified in the frequency
range of 4000 to 8000 $\mu$Hz down to an amplitude of 0.003 modulation intensity
level (Kilkenny et al. 2003). The sensitivity limits of the WET enabled the detection of very low-amplitude pulsations even during eclipse.  Fig. \ref{pg13} shows the light curve and frequency spectrum obtained during the campaign. The inclination of the system is estimated to be 81 degrees, and therefore even during primary eclipse, pulsations were expected. One important result obtained from the WET campaign was that during the eclipse, pulsations were indeed observed, however due to insufficient coverage of eclipses during the campaign the pulsational modes could not identified.

\subsection{WET Campaign on roAp Star HR\,1217}

The birth of the asteroseismology at ARIES began with the ``Nainital-Cape Survey'' project for the quest of new roAp stars in the Northern hemisphere. ARIES participated in the the WET campaign for HR\,1217, one of the best-studied roAp star which exhibited six oscillation frequencies each of these showing rotational modulation (Kurtz et al. 1989).  The result  showed that  5 adjacent frequencies groups in the previous data could be explained by splitting main frequency and the $6^{th}$ frequency is 3/4 of the splitting frequency was unexplained. To explain it,  a whole earth campaign was organized and data was collected over 35 d with a 34\% duty cycle in November - December 2000 when a total of 342 hr data through Johnson $B$ data with 10-s time resolution was obtained.  The precision of the derived amplitudes is 14-$\mu$mag, makes the highest precision ground-based photometric studies ever undertaken. The 
multi-site data clearly shows amplitude modulation for some modes between 1986
and 2000. 

\begin{figure}
\center{
\includegraphics[width=1.0\textwidth,height=0.2\textheight]{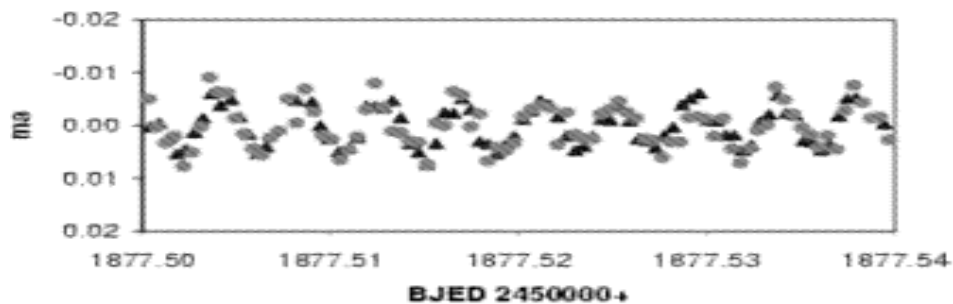}
}
\caption{: A section of 1-hr light curve of HR\,1217. Adopted from Kurtz et al. (2005).}
\label{hr1217-lc}
\end{figure}

\begin{figure}
\center{
\includegraphics[width=0.7\textwidth,height=0.7\textheight]{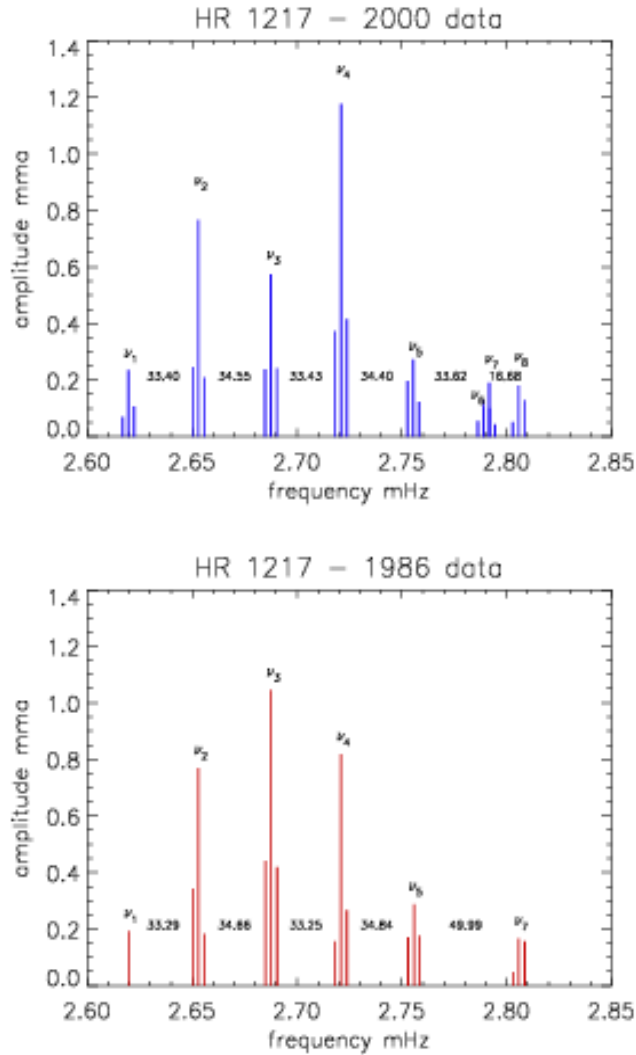}
}
\caption{: The difference in the Fourier spectra of the WET data set (top) and the 1986 dataset
(bottom). Adopted from Kurtz et al. (2002).}
\label{hr1217-ft}
\end{figure}

 Fig. \ref{hr1217-lc} shows a section of the light curve of HR\,1217 observed under this campaign and Fig. \ref{hr1217-ft} shows a schematic amplitude spectrum of the determined frequency. An additional peak was detected at 2788.95 $\mu$Hz with amplitude 0.1 mmag (Kurtz et al. 2002). The amplitudes of various modes change with time and this could be the reason for the non-detection of the new frequency in the 1986 dataset and its detection in 2000 can  be confirmed only with future observations. 

\subsection{WET Campaign on Pulsating White Dwarfs}

ARIES participated in the WET campaigns organized for  DA (ZZ Ceti) and DB white dwarfs. 
To check the stability of the pulsation periods an extensive multi-site data of ZZ Ceti was carried out. The analysis of these observations concluded that the characteristic stability time-scale for the pulsation period is more than 1.2 Gyr comparable to the theoretical cooling time-scale typical for this star (Mukadam et al. 2003).

\begin{figure}
\center{
\includegraphics[width=0.8\textwidth,height=0.4\textheight]{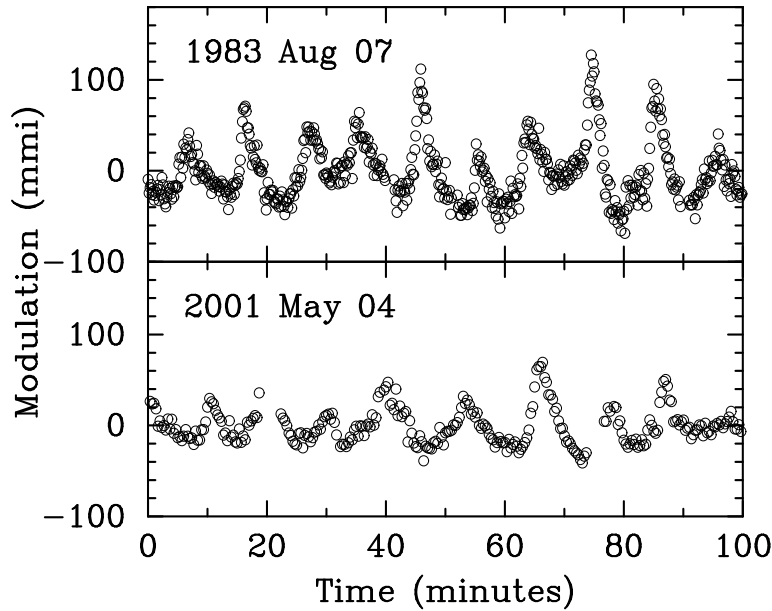}
}
\caption{: Upper panel: the discovery light curve of KUV05134+2605. Lower panel: section of one of the light curve acquired during the WET run. Note the change in the pulsational
time scales and amplitudes. Adopted from Handler (2003).}
\label{wdwarf}
\end{figure}

To study the temporal behavior of the pulsational amplitudes and frequencies of DBV pulsating  the time-series photometry of KUV05134+2605 were carried out under the WET campaign. Fig. \ref{wdwarf} shows a part of the light curve of KUV05134+2605 observed during this campaign. For the comparison, the  previously obtained light curve is also shown in the upper panel. The beating of multiple pulsation modes cannot explain simply and the amplitude variability could be intrinsic (Handler 2003).

\section{Future Prospects}
 
We have seen that many efforts from the ground and spaced based instruments are being taken to obtained the high-precision data to improve the  stellar models. However, there is a continuous  need to improve the observational situation and strong prospects for the stellar structure and evolution. To reduce the gaps in the data, ground based observations can be  carried out in a co-ordinated fashion by involving two or more observatories. This is one of the motivation for the development of the Stellar Observations Network Group (SONG)(Grundahl et al. 2009, 2011) which is consist of 8 nodes with a suitable geographical distribution in the Northern and Southern hemisphere. Each node has a robotic telescope of diameter 1-m equipped with a high-resolution spectrograph (resolution of 10$^5$) capable of reaching a RV precision of 1 $ms^{-1}$ for stars down to magnitude V = 6 for Doppler velocity observations and high-speed camera for photometry of crowded fields. The first node of SONG network is commissioning in Tenerife, a Chinese node is under construction, and additional nodes are expected to be added in the near future. 

The success of the MOST satellite gave the idea of building small and relatively inexpensive satellites (also called nano-satellites) the first space mission dedicated to the observations of the stellar oscillations (Walker et al. 2003). The BRIght Target Explorer mission (BRITE) is a constellation of  Austrian-Polish-Canadian mission of weight 7-kg was launched by Indian rocket PSLV-C20 on 25 February 2013 at  800-km high polar orbits. The  mission design consists of six nanosats (hence Constellation): two from Austria, two from Canada, and two from Poland. Each 7 kg nanosat carries an optical telescope of aperture 3-cm feeding an uncooled CCD. One instrument in each pair is equipped with a blue filter; the other with a red filter. Each BRITE instrument has a wide field of view ($\approx$24 degrees), so up to about 15 bright stars can be observed simultaneously, sampled in 32 pixels x 32 pixels sub-rasters (Kuschnig \& Weiss 2009). 
\begin{figure}
\center{
\includegraphics[width=0.8\textwidth,height=0.6\textheight]{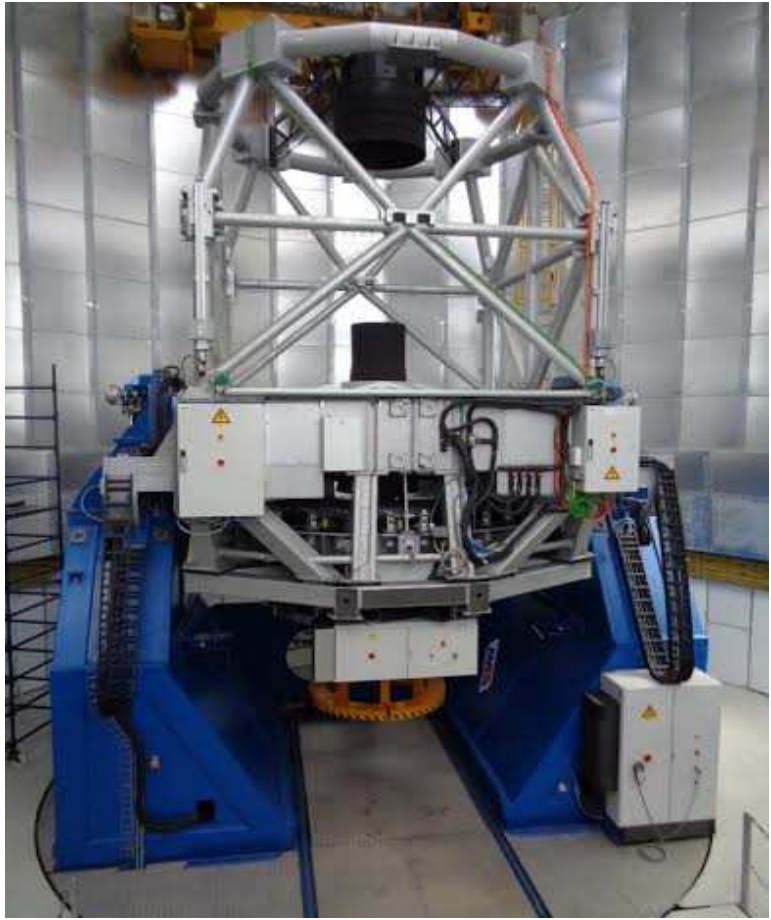}
}
\caption{: A picture of fully assembled 3.6-m telescope at Devasthal site.}
\label{3.6}
\end{figure}

Looking even farther to the future an European Space Agency ``Cosmic Vision
2015--2025'' proposal called PLAnetary Transits and Oscillations of
stars (PLATO) has been selected for ESA’s M3  expected to launch in 2022-2024 to providing accurate key planet parameters (radius, mass, density and age) in statistical numbers
it addresses fundamental questions such as: How do planetary systems form and evolve ? The PLATO 2.0 instrument consists of 34 small aperture telescopes (32 with 25-sec readout cadence and 2 with 2.5- sec cadence) providing a wide field-of-view (22x32 $deg^2$) and a large photometric magnitude range (4-16 mag). Asteroseismology will be performed for these bright stars to obtain highly accurate stellar parameters, including masses and ages (Rauer 2013). 

\section{3.6-m Optical Telescope at Devasthal}

The photon and scintillation are two dominating source of noise that play important role in the detection of the low-amplitude pulsational variability. Both of these noise can be minimized by observing with a bigger telescopes located at a good observing site. To reduce the scintillation noise for the detection of the low-amplitude light variations, a 3.6-m telescope at Devasthal site (longitude: $ 79^{o}40^{'}57{''} $ E, latitude : $29^{o}22{'}26{'}' $ N, altitude : 2420-m) has been installed and the first light is expected to be seen by mid of year 2015. Fig. \ref{3.6} shows a picture of fully assembled 3.6-m telescope at Devasthal. In the following subsections, we discuss the back-end instruments to be developed at ARIES for the asteroseismic study of pulsating variables. 

\subsection{Time-Series Photometers} 
\label{photo}

For the asteroseismic study of the transient phenomena we are developing a single-channel (focal reducer) and three-channel high-speed time-series CCD  photometer for the 3.6-m telescope. The scientific objective behind the building of these instruments is to perform ateroseismic study of transient events such as CP stars, white dwarfs, cataclysmic variables, red giants, open star clusters, detection of the extra-solar planetary systems. 

\subsection{Optical Design of the Time-series Photometers}

The starting point of optical design of  an instrument is to decide the required field of view (FoV) which should be large enough to have two- to three-comparison stars of comparable brightness and color to the target star. The probability of finding a comparison star of a given magnitude depends upon the search radius and the galactic latitude of the star (Simons et al. 1995). For example the probability of finding a comparison star for a 12 mag star at galactic latitude of 30$^o$ is 80 per cent if the search radius of 5 arcmin. Most of the target sources to be observed from 3.6-m telescope are expected fainter than 12 magnitude so a field of view of 6 arcmin virtually guarantees the presence of a suitable comparison stars. 
 
\subsubsection{Single-Channel Photometer} 

\begin{figure}
\center{
\includegraphics[width=1.1\textwidth,height=0.2\textheight]{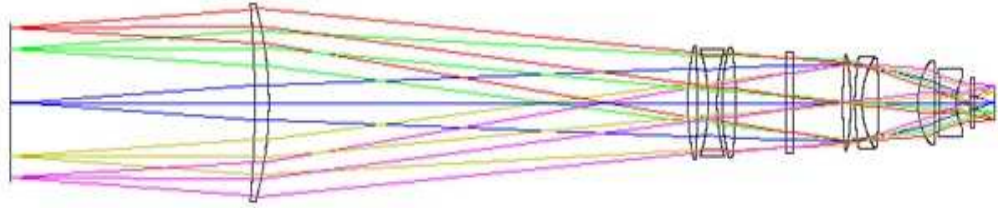}
}
\caption{: The optical design of the single-channel CCD photometer. }
\label{single}
\end{figure}

Fig. \ref{single} shows the preliminary optical design of a single-channel photometer for the side port of 3.6-m telescope. A total of ten optical elements are used to reduce the focal length to produce required FoV. The design was driven using an exit-pupil far behind the last collimator lens. This locates the pupil close or inside the cameras, simplifying the camera design and reducing the size of camera lenses. The collimator has not been designed to avoid ghost-image issues that would result from a non-parallel beam passing through the thick filters/dichroics. Optimization was done on the complete telescope--collimator--camera combination without varying any parameters of the telescope. This allowed us to benefit from additional degrees of freedom without having the image quality of the collimator. Collimator contains four lenses and camera contains five lenses. The last lens is a field flattener that corrects the combined field curvature from telescope, collimator and camera. The field lens also acts as the vacuum seal of the cryostat. This avoids the need for an additional plane vacuum window and thus reduces Fresnel reflections. The collimator and camera are optimized combinedly with filter and detector window.
 
\subsubsection{Three-Channel Focal Reducer} 

\begin{figure}
\center{
\includegraphics[width=1.1\textwidth,height=0.4\textheight]{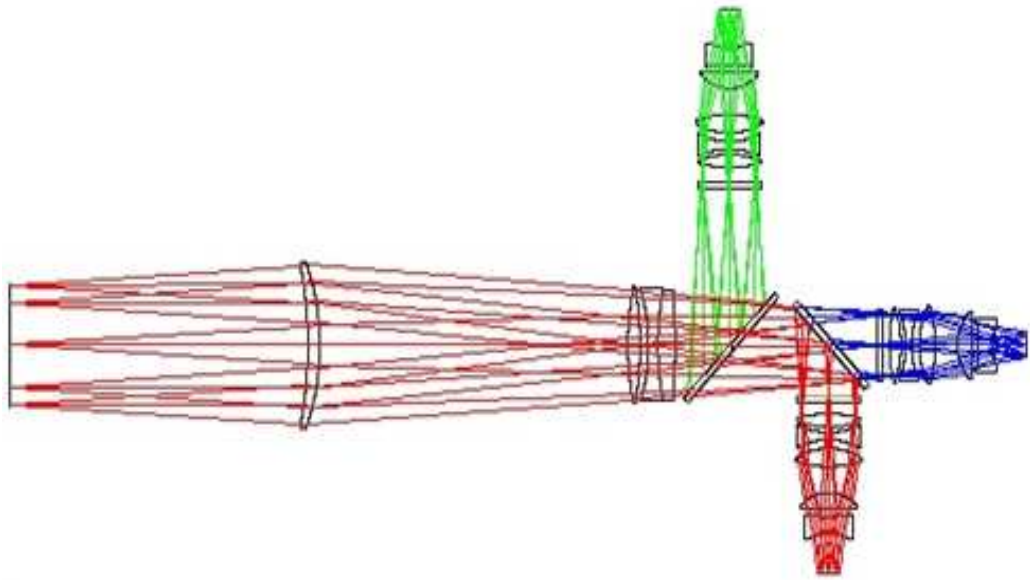}
}
\caption{: The optical design of the proposed three-channel CCD photometer.}
\label{three}
\end{figure}

A single-channel photometer can be used to observe stars in a single band. For  multi-band photometric observations this gives poor time-resolution. Hence we decided to develop a three-channel fast CCD photometer for observations of pulsating variables in three colors simultaneously.  The optical design of the three-channel photometer is modified from the single-channel photometer by inserting fused silica dichroic beam splitters and N-BK7 SDSS filters. Fig. \ref{three} shows the preliminary optical design of the three-channel photometer  made for the axial port of 3.6-m telescope. The design consists of a four element collimator and three same set of five camera lenses collecting the beam from dichroic beam splitters and send it to the detector through filters. 

\subsection{Detectors}

The choice of detector for these instruments is frame transfer based CCD. To have the high-quantum efficiency (more than 80\%) in the wavelength range 4500 \AA \,- 7500 \AA \,, we prefer thin and back illuminated CCD. In the market the maximum no. of pixels in such CCDs are 1024 x 1024 each of size 13 $\mu$ a side.  The chip will be cooled by a three-stage thermo-electric cooler (TEC) based on the Peltier effect.

\subsection{Filters}

For the high-speed time-series photometry we plan to use both the standard SDSS $u,g,r,i,z$ and Johnson Cousins U,B,V,R$_c$,I$_c$ filters. Provision for the broad band filter BG40 made of Scott glass would be useful for the asteroseismic observations of pulsating white dwarfs.
 
\subsection{System Throughput}

The performance of an instrument depends upon the optimization of its throughput.
Therefore to observe the objects of a wide range of brightness one would like to know the Signal-to-Noise (S/N) ratio achievable for a given set of exposure times. Given the transmission coefficients of the mirrors, filters, CCD glass, brightness of sky, extinction, quantum efficiency of the CCD chip, the S/N ratio can be calculated  using the formula given by  Mayya et al. (1991). If the sky brightness and the CCD input parameters like pixel size, dark current, read out noise are known then one can calculate the sky counts, the underlying S/N ratio and photometric accuracy (McLean 1989).

\begin{figure}
\center
\includegraphics[width=0.9\textwidth,height=0.3\textheight]{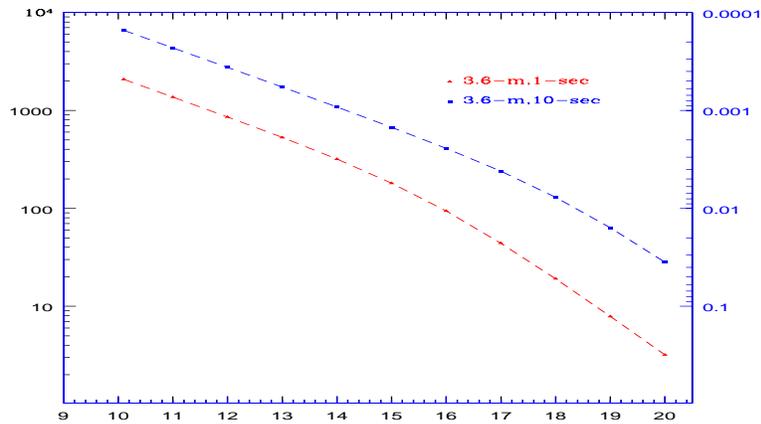}
\caption{: A plot of B-band magnitude versus the calculated signal-to-noise  ratio (y-axis, left) and corresponding error in the magnitude determinations (y-axis,right) for exposure times 1-sec and 10-sec for single-channel time-series CCD photometer for the 3.6-m telescope.}
\label{thB}
\end{figure}

\begin{figure}
\center
\includegraphics[width=0.9\textwidth,height=0.3\textheight]{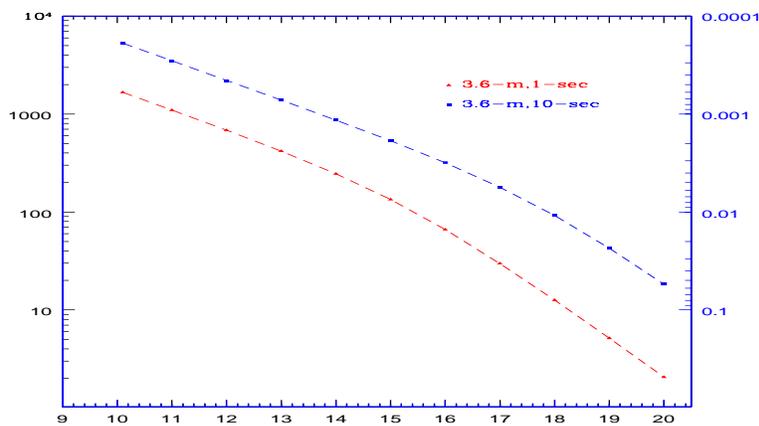}
\caption{: Similar to Fig. \ref{thB} but for V-band.}
\label{thV}
\end{figure}

We have calculated the system throughput for the single-channel photometer to be equipped at the side port of 3.6-m telescope. Fig. \ref{thB} and \ref{thV}  show the B and V-band detection limit of the 3.6-m telescope for the stars of magnitude range 10 to 20 with exposure time 1 and 10-sec. From this figure it is evident that  for the 3.6-m telescope the photometric precision of 0.01 mag in B-band can be achieved for stars  of 17.5 mag and 18.5 mag with same exposure time.  These figures also show the detection limit in the B and V-band, respectively with the same exposure time. 

\section{Conclusion}

Asteroseismology is an unique approach for the investigation of stellar structure and evolution which has significantly improved our knowledge in last decades. Though asteroseismology is still in its infancy and far from the helioseismology however the high-precision ground and space based data  will allow us to make the seismic analysis so robust that one can obtain the reliable information on the internal structure of stars. It is true that the space based time-series data are better than the ground-based in terms of accuracy and higher duty cycle but the ground based observations are more flexible in the selection of targets, time and duration of observations. The successful asteroseismology programme looks set to continue in the re-purposed Kepler Mission, K2 (Chaplin et al. 2013; Howell et al. 2014). Observationally, the longitude of India makes this region critical to all global observational helioseismic and asteroseismic studies to fill the gap in the time-series data. The presence of 4-m class optical telescope in the Asian region  would be very useful for asteroseismic study of faint pulsating variables. At ARIES we are developing a single and three-channel fast CCD  photometers to optimized for the high-speed time-series measurements and looking forward for their commissioning. In the future, the asteroseismology of many thousands of stars in various stages of their evolution will enable us to tackle several long-standing problems, such as stellar dynamos, stellar convection.


\section*{Acknowledgments}
SJ dedicates this review article to his mother Late Smt. Ganga Joshi. SJ is grateful to the SOC of the Indo-UK seminar for the invitation to deliver an invited talk on the topic ``Asteroseismology from ARIES''. We acknowledge Prof. Ram Sagar, Prof. D. W. Kurtz, Dr. Peter Martinez, Dr. S. Seetha and Dr. B. N. Ashoka for initiating the field of Asteroseismology at ARIES. The authors are thankful to the anonymous referee for providing the remarkable comments that improved the manuscript drastically. SJ acknowledge to Sowgata Chaudhary for reading the manuscript carefully and pointing out the typos. The optical design of the time-series photometer has been done by Er. Krishna Reddy. This is the compilation of work done under the Indo-Russian RFBR project INT/RFBR/P-118.

\end{document}